\RequirePackage{lineno}
\documentclass[prd,twocolumn,showpacs,superscriptaddress,amsmath,amssymb]{revtex4-1}
\usepackage{graphicx}
\usepackage{epsfig}
\usepackage{dcolumn}
\usepackage{bm}
\usepackage{overpic}
\usepackage{color}
\usepackage{lineno}
\usepackage{subfigure}
\usepackage{ulem}
\usepackage{url}
\include{def-com}
\def \jpsi {J/\psi}
\def \psip {\psi(3686)}
\def \epem {e^{+}e^{-}}
\def \piz  {\pi^0}
\def \pzpz {\pi^0\pi^0}
\def \pppm {\pi^{+}\pi^{-}}
\def \pipm {\pi^{\pm}}

\def \lplm {\ell^+\ell^-}
\def\pip{\pi^{+}}
\def\pim{\pi^{-}}

\def \zcbs {Z_c(3900)^\pm}

\def \gev  {\mbox{GeV}}
\def \gevc {\mbox{GeV/$c$}}
\def \gevcc{\mbox{GeV/$c^2$}}
\def \mev  {\mbox{MeV}}
\def \mevcc{\mbox{MeV/$c^2$}}
\def \ifb  {\mbox{fb$^{-1}$}}
\begin{document}
\title{\boldmath Measurement of $\epem\to\pppm\psip$ from 4.008 to 4.600 $\gev$ and
                 observation of a charged structure in the $\pi^\pm\psip$ mass spectrum}
\author{
M.~Ablikim$^{1}$, M.~N.~Achasov$^{9,d}$, S. ~Ahmed$^{14}$, X.~C.~Ai$^{1}$, O.~Albayrak$^{5}$, M.~Albrecht$^{4}$, D.~J.~Ambrose$^{45}$, A.~Amoroso$^{50A,50C}$, F.~F.~An$^{1}$, Q.~An$^{47,38}$
% ...
\\
\vspace{0.2cm}
(BESIII Collaboration)\\
\vspace{0.2cm} {\it
$^{1}$ Institute of High Energy Physics, Beijing 100049, People's Republic of China\\
% ...
$^{4}$ Bochum Ruhr-University, D-44780 Bochum, Germany\\
$^{5}$ Carnegie Mellon University, Pittsburgh, Pennsylvania 15213, USA\\
% ...
$^{9}$ G.I. Budker Institute of Nuclear Physics SB RAS (BINP), Novosibirsk 630090, Russia\\
% ...
$^{14}$ Helmholtz Institute Mainz, Johann-Joachim-Becher-Weg 45, D-55099 Mainz, Germany\\
% ...
$^{38}$ State Key Laboratory of Particle Detection and Electronics, Beijing 100049, Hefei 230026, People's Republic of China\\
% ...
$^{45}$ University of Rochester, Rochester, New York 14627, USA\\
% ...
$^{47}$ University of Science and Technology of China, Hefei 230026, People's Republic of China\\
% ...
$^{50}$ (A)University of Turin, I-10125, Turin, Italy; (B)University of Eastern Piedmont, I-15121, Alessandria, Italy; (C)INFN, I-10125, Turin, Italy\\
% ...
$^{d}$ Also at the Novosibirsk State University, Novosibirsk, 630090, Russia\\
% ...
 }
\vspace{0.4cm}
}

\date{\today}

\begin{abstract}
We study the process $\epem\to\pppm\psip$ using 5.1 $\ifb$ of data collected at 16 center-of-mass energy ($\sqrt{s}$) points from 4.008 to 4.600~$\gev$ by the BESIII detector operating at the BEPCII collider.
The measured Born cross sections for $\epem\to\pppm\psip$ are consistent with previous results, but with much improved precision.
A fit to the cross section shows contributions from two structures:
the first has $M=4209.5\pm7.4\pm1.4$~MeV/$c^{2}$ and $\Gamma=80.1\pm24.6\pm2.9$~MeV,
and the second has
$M=4383.8\pm4.2\pm0.8$~MeV/c$^{2}$ and $\Gamma=84.2\pm12.5\pm2.1$~MeV,
where the first errors are statistical and the second systematic.
The lower-mass resonance
is observed in the process $\epem\to\pppm\psip$ for the first time with a statistical significance of $5.8\sigma$.
A charged charmonium-like structure is observed in the $\pipm\psip$ invariant mass spectrum
for data at $\sqrt{s} = 4.416~\gev$.
A fit with an $S$-wave Breit-Wigner function
yields a mass $M=4032.1\pm2.4$~$\mevcc$,
where the errors are statistical only.
However, there are still unresolved discrepancies between the fit model and data.
The width of the intermediate state varies in a wide range for
different kinematic regions within the data set.
Therefore no simple interpretation of the data has been found,
and a future data sample with larger statistics and more theoretical input will be required to better understand this issue.

\end{abstract}


\maketitle
\section{\boldmath INTRODUCTION}
In the past decade, a series of new vector ($J^{PC}= 1^{--}$) charmoniumlike states, {\it e.g.}, the $Y(4260)$, $Y(4360)$, and $Y(4660)$, have been observed in $\epem$ annihilation through dipion hadronic transitions to low mass charmonium states~\cite{Y4260BaBar,Y4260Cleo, Y4260Belle, Y4360Belle, Y4360BaBar}.
Many theoretical interpretations have been proposed to understand the underlying structure of the $Y$ family of states, such as hybrid charmonium~\cite{Y4360hybrid}, tetraquark~\cite{Y4360tetra}, or hadronic molecule~\cite{Y4360molecule} {\it etc}.
The $Y(4360)$ was first observed in $\epem\to\gamma_{\rm ISR}Y(4360)\to\gamma_{\rm ISR}\pppm\psip$ by BaBar~\cite{Y4360BaBar} and subsequently confirmed by Belle~\cite{Y4360Belle}.
Recently, two resonant structures were
observed in the processes $e^{+}e^{-}\to\pi^{+}\pi^{-}J/\psi$ at BESIII~\cite{pipijpsi},
indicating the $Y(4260)$ resonance reported by previous experiments~\cite{Y4260BaBar,Y4260Cleo, Y4260Belle}
actually consists of two structures.
There are two resonances observed in the process $e^{+}e^{-}\to\pi^{+}\pi^{-}h_{c}$~\cite{pipihc}
in the mass region between 4.2 and 4.4~GeV/$c^{2}$.
The lower mass state, $Y(4220)$, has a mass consistent
with the lower mass state observed in $e^{+}e^{-}\to\pi^{+}\pi^{-}J/\psi$~\cite{pipijpsi},
but has a somewhat larger width. The higher mass state, $Y(4390)$,
does not match any known vector charmonium or charmonium-like
states.
Therefore, a more precise measurement of the cross section of  $\epem\to\pppm\psip$ at BESIII will help to
clarify the spectrum of vector particles.

In recent years, another mysterious and interesting new pattern of charmoniumlike states, the $Z_c^{\pm}$'s, have been observed both in final states containing a charged pion and a low mass charmonium state~\cite{Y4260Belle, ZC3900BES, ZC3900CLEO, ZC4020BES}, and in pairs of charmed mesons $(D^{(\ast)}\bar{D}^{\ast})^\pm$~\cite{ZC3885BES,ZC3885BES2,ZC4025BES}.
A similar pattern of states is also observed in the bottomonium system~\cite{zb}.
All of these states contain a minimum of four quarks as they are electrically charged and their constituents include a $c\bar{c}$ pair.
They are good candidates for meson-meson molecules or tightly bound tetraquark states.
To clarify their true nature, it is essential  to search for this kind of exotic state in other final states.
Belle's recently updated result of $\epem\to\gamma_{\rm ISR}\pppm\psip$, using their full data samples, shows evidence for a new $Z_c^\pm$ state in the mass spectrum of $\pi^{\pm}\psip$  with a mass around $4.05$ $\gevcc$ and a statistical significance of $3.5\sigma$~\cite{Y4360Belle}.
It is important to confirm the existence of this new $Z_c^{\pm}$ candidate in other experiments.

In this paper, we present a study of $\epem\to\pip\pim\psip$, with two decay modes $\psip\to\pppm\jpsi$ (mode I) and $\psip\to \text{neutrals}+\jpsi$ (mode II), where `neutrals' refers to $\pzpz$, $\piz$, $\eta$ and $\gamma\gamma$, at center-of-mass (c.m.)~energies $\sqrt{s}$ from 4.008 to 4.600~$\gev$. The $\jpsi$ is reconstructed in its prominent decay mode $\jpsi\to\lplm$ ($\ell=e/\mu$).
The actual c.m.~energies are measured by studying the process $e^+e^-\to \mu^+\mu^-$~\cite{cme}.
The data samples used in this analysis were collected with the BESIII detector at the BEPCII collider at 16 different c.m.~energy points with a total integrated luminosity of 5.1~$\ifb$~\cite{luminosity}.

\section{\boldmath THE BESIII EXPERIMENT AND THE DATA SETS}
BEPCII is a double-ring $e^{+}e^{-}$ collider running at c.m.~energies between 2.0 and 4.6~GeV,
reaching a peak luminosity of $1.0\times10^{33}$cm$^{-2}$s$^{-1}$ at a c.m.~energy
of 3770~MeV.
The cylindrical BESIII detector has an effective geometrical acceptance of $93\%$ of 4$\pi$
and is divided into a barrel section and two endcaps.
It contains a small cell, helium-based (60$\%$ He, 40$\%$ C$_{3}$H$_{8}$) main drift chamber
(MDC) which provides a momentum measurement of charged particles with a resolution of $0.5\%$
at a momentum of 1~GeV/c in a magnetic field of 1 Tesla.
The energy loss measurement ($dE/dx$) provided by the MDC has a resolution better than $6\%$.
A time-of-flight system (TOF) consisting of 5-cm-thick plastic scintillators can measure the
flight time of charged particles with a time resolution of 80 ps in the barrel and 110 ps in
the endcaps.
An electromagnetic calorimeter (EMC) consisting of 6240 CsI~(Tl) in a cylindrical structure
and two endcaps is used to measure the energies of photons and electrons.
The energy resolution of the EMC is $2.5\%$ in the barrel and $5.0\%$ in the endcaps for
photon/electrons with an energy of 1 GeV.
The position resolution of the EMC is 6~mm in the barrel and 9~mm in the endcaps. A muon
system (MUC) consisting of 1000~m$^{2}$ of Resistive Plate Chambers (RPC) is used to identify
muons and provides a spatial resolution better than 2~cm.
A detailed description of the BESIII detector can be found in Ref.~\cite{besint}.

The {\sc Geant4}-based~\cite{geant4} Monte Carlo (MC) simulation software package {\sc Boost}~\cite{Deng} is used to generate the signal and background MC samples.
The $\epem$ collision is simulated with the {\sc KKMC}~\cite{kkmc} generator taking into account the beam energy spread and Initial State Radiation (ISR), where the line shape of the $\epem\to\pppm\psip$ cross section is taken from the latest Belle results~\cite{Y4360Belle}.
The QED Final State Radiative (FSR) correction for $\jpsi\to\lplm$ is incorporated with \textsc{Photos}~\cite{photos}.
Large signal MC samples $\epem\to\pppm\psip$ as well as the dominant backgrounds are generated exclusively at each c.m.~energy, where the decay $\psip\to \pi^{+(0)}\pi^{-(0)} \jpsi$ is simulated with the \textsc{Jpipi} model~\cite{jpipi}, and the decay $\epem\to\pppm\psip$ is generated with both the \textsc{Jpipi} model and a model distributed uniformly in phase space~(PHSP).
The {\sc Jpipi} model is constructed with an effective Lagrangian for
the hadronic decays and the coupling constants in the model are obtained by
a fit to the invariant $\pi^{+}\pi^{-}$ mass distributions in
$\psi(3686)\to\pi^{+}\pi^{-}J/\psi$.
The generic inclusive MC samples at $\sqrt{s}= 4.258$ and 4.358~$\gev$, with integrated luminosity equivalent to that of data, are
generated to study the potential background contributions, where the
known decay modes are generated with {\sc Evtgen} \cite{evtgen}
with branching fractions set to values in the Particle Data Group
(PDG)~\cite{pdg},
and the remaining unknown ones are generated with \textsc{Lundcharm} \cite{lundcharm}.

\section{\boldmath Event selection}

The final states of the decay $\epem\to\pppm\psip$ have topologies $\pppm\pppm\lplm$ and $\pppm$$\lplm+N\gamma$~($N\geq2)$ for mode I and II, respectively.
A good charged track must have a polar angle
$|\cos\theta|<0.93$ in the MDC, and have the point of closest approach to the interaction
point within 1~cm in the plane perpendicular to the beam and within $\pm10$~cm
along the beam.
Candidate events are required to have five (with $\pm1$ net charge) or six (with zero net charge) charged tracks for mode I and four (with zero net charge)  charged tracks for mode II.
Photon candidates are reconstructed by clustering EMC energy deposits.
They must have a minimum energy of 25~$\mev$ for $|\cos\theta|<$~0.8
or 50~$\mev$ for 0.86~$<|\cos\theta|<$~0.92,
and be isolated from all charged tracks by an angle $>10^{\circ}$.
EMC timing is used
to suppress electronic noise and energy deposition unrelated to the event.
At least two  photon candidates are required for mode II.
Since pions and leptons are kinematically well separated for the signal process, charged tracks with momentum $p>1.0$~$\gevc$ are assumed to be leptons, while those with $p< 0.65 \;\gevc$ are pions.
Electron and muon separation is performed with the deposited energy $E$ in the electromagnetic calorimeter.
Electrons must have $E/pc> 0.7$, while muons have $E < 0.45 \;\gev$.
Candidate events are required to have one lepton pair $\lplm$, and the other tracks must be pions.
The number of hits in the muon counter is further required to be not less than 5 for $\jpsi\to\mu^+\mu^-$ candidates.

In mode I, a four-constraint (4C) kinematic fit imposing energy-momentum conservation under the $\epem\to\pppm\pppm\lplm$ hypothesis is applied for the event candidates with six charged tracks, and a one-constraint (1C) kinematic fit with one missing pion is applied for the candidates with five charged tracks.
The corresponding $\chi^2$ of the fit must be $\chi^{2}_{4C}<60$ or $\chi^{2}_{1C}<15$, respectively.
The $\jpsi$ signal appears with low background in the spectrum of the $\lplm$ invariant mass, $M(\lplm)$, calculated with the corrected momentum after the kinematic fit, and is required to be within $3.05 < M(\lplm) < 3.15 \;\gevcc$.
To improve the mass resolution of the $\psip$, the selected candidates with six or five good charged tracks are further fed into a five-constraint (5C) or two-constraint (2C) kinematic fit with an additional constraint on $M(\lplm)$ to the nominal $\jpsi$ mass ($M(\jpsi)$)~\cite{pdg}, respectively.
The $\psip$ signal is reconstructed with the $\pppm\jpsi$ system (four combinations per event) whose invariant mass $M(\pppm\jpsi)$, calculated with the corrected momentum after the 5C or 2C fit, is closest to the nominal $\psip$ mass ($M(\psip)$)~\cite{pdg}.

In mode II, no kinematic fit is applied.
The $\jpsi$ signal is extracted by requiring $3.05 < M(\lplm) < 3.15 \;\gevcc$.
The $\psip$ signal is reconstructed using the mass recoiling against the $\pppm$ system, $M^{\rm recoil}(\pppm)$.
The requirement $|M^{\rm corr}_{\psip}-M(\psip)|> 8 \;\mevcc$ is imposed to veto the background $\epem\to \text{neutrals} + \psip$  ($\psip\to\pppm\jpsi$), where $M^{\rm corr}_{\psip}$ = $M(\pppm\lplm)-M(\lplm)+M(\jpsi)$.
A requirement $|M(\gamma\gamma\pppm)-M(\eta)|> 50 \;\mevcc$ is used to eliminate the background $\epem\to\eta\jpsi$ ($\eta\to\pppm\piz$), where $\gamma\gamma$ are the two photons with largest energy, and $M(\eta)$ is the nominal $\eta$ mass~\cite{pdg}.
A requirement $\cos\theta_{\pppm} < 0.9$ is also applied to remove the radiative Bhabha and dimuon background in which a photon converts into a $\epem$ pair and is misidentified as a $\pppm$ pair.

\section{\boldmath { Extraction of the Born cross section}}

With the above selection criteria, a prominent $\psip$ signal over a small background is observed in the $M(\pppm\jpsi)$ and $M^{\rm recoil}(\pppm)$ spectra for Mode I and Mode II, respectively, as shown in Fig.~\ref{mass} for data at $\sqrt{s}= 4.416\;\gev$ as an example.
Potential backgrounds without a $\jpsi$ in the final state are explored with events within the $\jpsi$ sideband regions.
Backgrounds including a $\jpsi$ in the final state are studied with both inclusive and exclusive MC samples.
From these studies, no peaking background is expected in the range of interest in the two $\psip$ decay modes.
To determine the signal yields, unbinned maximum likelihood fits are performed to the $M(\pppm\jpsi)$ and $M^{\rm recoil}(\pppm)$ spectra for the two modes, respectively.
Here, the signal probability density function (PDF) is described with the MC simulated shape convoluted with a Gaussian function to account for the mass resolution difference between data and MC simulation, and the PDF for background is presented as a linear function.
Simultaneous fits to data at 16 different c.m.~energies are performed.  The c.m.~energies share a common background shape and a common data/MC resolution difference.

\begin{figure}[htbp]
\begin{center}
\setlength{\belowcaptionskip}{-0.6cm}
\begin{overpic}[width=4.2cm,height=3.8cm,angle=0]{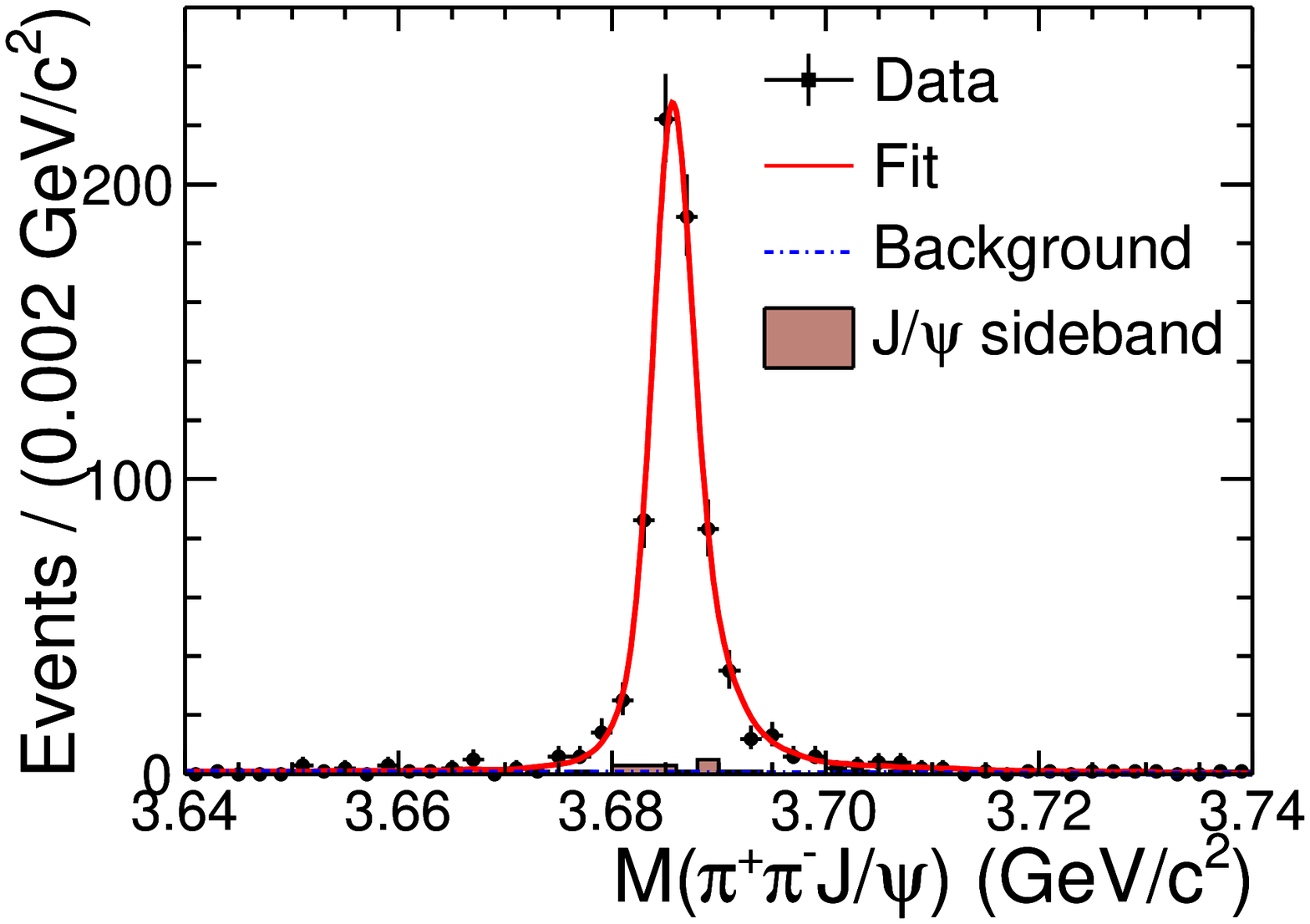}
\end{overpic}
\begin{overpic}[width=4.2cm,height=3.8cm,angle=0]{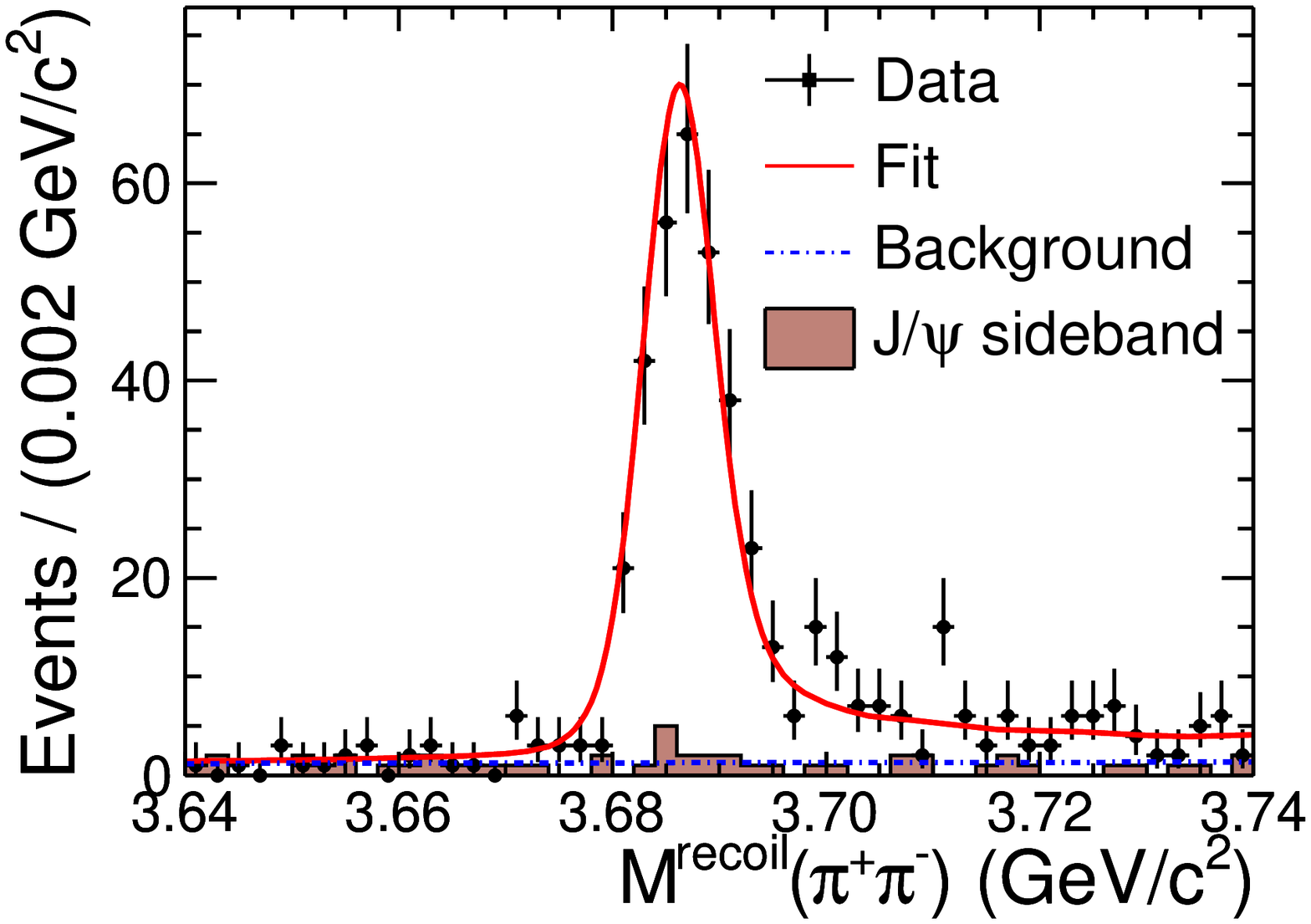}
\end{overpic}
\caption{ Distributions of (left) $M(\pppm\jpsi)$ for mode I and (right) $M^{\rm recoil}(\pppm)$ for mode II at $\sqrt{s}=$4.416~GeV,
with the integrated luminosity of 1074~pb$^{-1}$.
         Dots with error bars are data, the shaded histograms are background from $\jpsi$ sideband, and the curves are the fit described in the text. }
\label{mass}
\end{center}
\end{figure}

The Born cross section is determined from
\begin{equation}
\sigma^{\rm B}_{i} = \frac{N^{\rm obs}_{i}}
            {\mathcal{L}_{\rm int}~(1+ \delta^{r})~(1+ \delta^{v})~\mathcal{B}r_{i}~\epsilon_{i}}~(i=1,~2),
\label{equ:born}
\end{equation}
where
$N^{\rm obs}_i$ is the signal yield extracted from the fit,
$\mathcal{L}_{\rm int}$ is the integrated luminosity,
$1+\delta^{r}$ is the ISR correction factor,
$1+\delta^{v}$ is the vacuum polarization factor,
$\mathcal{B}r$ is the product of the decay branching fractions of the intermediate states
in the cascade decay (4.11\% for mode I and 2.95\% for mode II), and $\epsilon$ is the detection efficiency.
The various numbers used in the Born cross section calculation are summarized in Table~\ref{result1}.

\begin{table*}[htbp]
\setlength{\belowcaptionskip}{-0.1cm}
  \caption{
  Summary of the measurement of the Born cross section $\sigma^{B}$ at individual c.~m.~energies.
  The subscript $1$ or $2$ denotes mode I or II, respectively.
  The first uncertainties are statistical, and the second systematic.
  An upper limit at the 90\% confidence level (C.L.) is determined by a
  profile likelihood method~\cite{uplimit} for data samples with low signal significance.
  }\label{result1}
  \footnotesize
  \begin{center}
  \begin{tabular}{ c | c | c|c|c|c|c|c|c|c|c}
  \hline
  \hline
  $\sqrt{s}$~(GeV) & $L$~(pb$^{-1}$) & $N^\text{obs}_{1}$ & $\epsilon_{1}$~(\%) & $N^\text{obs}_{2}$ & $\epsilon_{2}~(\%)$ &  $(1+\delta^{r})$ & $(1+\delta^{v})$ & $\sigma^{B}_{1}$~(pb) & $\sigma^{B}_{2}$~(pb) & $\sigma^{B}$~(pb)\\
  \hline
  4.008 &  482 &  $0.0\pm0.6$  & 22.6  &  $0.2\pm2.3$ &  4.7 & 0.70 & 1.056 &  $<0.9$ & $<23.3$ & $<0.9$\\
  4.085 &   52.6 &  $4.0\pm2.0$  & 36.1  &  $1.0\pm1.0$ & 20.9 & 0.75 & 1.056 &  $6.5\pm3.2\pm0.9$ & $3.9\pm3.9\pm0.3$   & $5.4\pm2.5\pm0.6$\\
  4.189 &   43.1 &  $3.8\pm2.0$  & 39.2  &  $2.8\pm2.2$ & 27.7 & 0.76 & 1.056 &  $6.8\pm3.6\pm0.7$ & $9.9\pm7.8\pm1.5$   & $7.3\pm3.3\pm0.7$\\
  4.208 &   54.6 & $8.9\pm3.3$  & 40.7  &  $7.0\pm3.0$ & 27.5 & 0.76 & 1.057 & $12.2\pm4.5\pm1.7$ & $20.0\pm8.4\pm1.9$  & $14.0\pm4.0\pm1.5$\\
  4.217 &   54.1 & $13.0\pm3.6$  & 40.9  &  $0.0\pm0.7$ & 27.3 & 0.76 & 1.057 & $17.8\pm4.9\pm1.5$ & $<30.4$  & $17.8\pm4.9\pm1.5$\\
  4.226 & 1092 &   $315\pm18$  & 39.2  &   $141\pm14$ & 28.1 & 0.76 & 1.056 & $22.3\pm1.3\pm1.7$ & $19.4\pm1.9\pm2.0$  & $21.3\pm1.1\pm1.6$\\
  4.242 &   55.6 & $11.0\pm3.3$  & 41.4  &  $7.9\pm3.1$ & 28.0 & 0.76 & 1.053 & $14.6\pm4.4\pm1.3$ & $21.5\pm8.4\pm1.8$  & $16.0\pm3.9\pm1.2$\\
  4.258 &  826 &   $241\pm16$  & 40.3  &    $84\pm11$ & 23.5 & 0.76 & 1.054 & $22.0\pm1.4\pm1.7$ & $18.3\pm2.5\pm1.8$  & $20.9\pm1.2\pm1.5$\\
  4.308 &   44.9 & $17.0\pm4.2$  & 41.6  & $15.0\pm4.1$ & 27.3 & 0.74 & 1.053 & $28.2\pm6.9\pm2.6$ & $53.2\pm14.5\pm7.4$ & $32.1\pm6.2\pm2.8$\\
  4.358 &  540 &   $439\pm21$  & 41.2  &   $275\pm19$ & 29.8 & 0.79 & 1.051 & $57.8\pm2.8\pm4.4$ & $69.8\pm4.8\pm5.2$  & $61.0\pm2.4\pm4.3$\\
  4.387 &   55.2 & $56.6\pm7.6$  & 39.4  & $25.7\pm6.1$ & 29.4 & 0.86 & 1.051 & $70.1\pm9.4\pm6.7$ & $59.4\pm14.1\pm5.6$ & $66.4\pm7.8\pm5.5$\\
  4.416 & 1074 &   $693\pm27$  & 37.8  &   $415\pm24$ & 27.4 & 0.96 & 1.053 & $41.0\pm1.6\pm3.2$ & $47.3\pm2.7\pm3.7$  & $42.8\pm1.4\pm3.0$\\
  4.467 &  110 & $15.1\pm4.2$  & 32.6  & $8.3\pm4.2$ & 23.9 & 1.10 & 1.055 &  $8.8\pm2.5\pm1.0$ & $9.2\pm4.7\pm1.9$  & $8.9\pm2.2\pm0.9$\\
  4.527 &  110 & $13.4\pm4.0$  & 29.1  & $7.0\pm3.6$ & 20.8 & 1.25 & 1.055 &  $7.7\pm2.3\pm0.9$ & $7.8\pm4.0\pm1.2$  & $7.7\pm2.0\pm0.8$\\
  4.575 &   47.7 &  $4.5\pm2.3$  & 28.3  &  $5.7\pm3.2$ & 20.2 & 1.23 & 1.055 &  $6.2\pm3.2\pm0.8$ & $15.4\pm8.7\pm1.8$  & $7.3\pm3.0\pm0.8$\\
  4.600 &  567 &   $106\pm11$  & 31.8  &  $71\pm10$ & 21.5 & 1.08 & 1.055 & $12.6\pm1.3\pm1.2$ & $17.2\pm2.4\pm1.6$  & $14.6\pm1.1\pm1.1$\\
  \hline \hline
  \end{tabular}
  \end{center}
  \end{table*}

Several sources of systematic uncertainty are considered in the measurement of the Born cross section.
The uncertainty in the tracking efficiency for leptons and pions are 1.0\% per track~\cite{trkuncertainty}.
The uncertainty in the photon efficiency is 1.0\% per photon~\cite{phouncertainty}.
The uncertainty related to the requirement on the number of hits in the muon counter is 4.2\%, determined by studying a control sample of $\epem\to\pppm\jpsi$.
The uncertainty associated with the kinematic fit is estimated with the same technique as in Ref.~\cite{kinematic}.
The uncertainty associated with the $\jpsi$ mass requirement is estimated by smearing
the $M(\lplm)$ distribution of MC samples according to the resolution of $M(\lplm)$ in data, and the difference in MC efficiencies is taken as the uncertainty.
In mode II, the uncertainties associated with the requirements on $M^{corr}_{\psip}$, $M(\gamma\gamma\pppm)$ and $\cos\theta_{\pppm}$ are evaluated by changing the requirements.
The uncertainty associated with the fit procedure is investigated by varying the fit range, replacing the linear function by a second-order polynomial function for the background, and varying the width of the Gaussian function for signal.
For data sets with large luminosity, the detection efficiency is estimated with a MC sample weighted according to the observed Dalitz plot $M^2(\pip\psip)$ versus $M^2(\pppm)$.
The corresponding uncertainty is estimated by varying the weight factors according to the statistical uncertainty of data in each bin on the Dalitz plane.
For data sets with low luminosity, the detection efficiency is estimated using MC samples with the \textsc{Jpipi} model;
the corresponding uncertainty is evaluated with large luminosity data sets, by taking the largest difference of efficiencies between the \textsc{Jpipi} model MC sample and
their nominal values with weighted MC.
The uncertainty related with the ISR correction factor is studied by replacing the input cross section line shape with the latest results from BaBar~\cite{Y4360BaBar} in KKMC, and the change on $(1+ \delta^{v})~\epsilon$ is taken as an uncertainty.
The uncertainty in the vacuum polarization factor is 0.5\% taken from a QED calculation~\cite{vacuum}.
The uncertainty on the integrated luminosity is 1\%, determined with large angle Bhabha events~\cite{luminosity}.
The uncertainties in the decay branching fractions of intermediate states are quoted from the PDG~\cite{pdg}.
The uncertainty from others sources, such as lepton separation, trigger efficiency, and FSR are negligible, and are conservatively taken to be 1.0\%.
Assuming all sources of systematic are independent, the total uncertainties are obtained by adding the individual values in quadrature, and are in a range between $7.7\%$ to $14.1\%$ and $7.4\%$ to $20.1\%$, depending on c.m.~energy, for mode I and II, respectively.

The measured Born cross sections of $\epem\to\pppm$$\psip$ at individual c.m.~energies for the two $\psip$ decay modes are consistent with each other within their uncertainties.
The measurements are therefore combined by considering the correlated and un-correlated uncertainties between the two modes, according to Refs.~\cite{combine1, combine2}.
The comparison of the combined Born cross section of $\epem\to\pppm\psip$ with those from previous experimental results is shown in Fig.~\ref{comresult1}.
The results are consistent with former experiments, and have much improved precision.

\begin{figure}[htbp]
\begin{center}
\includegraphics[width=7.6cm, height=6.5cm]{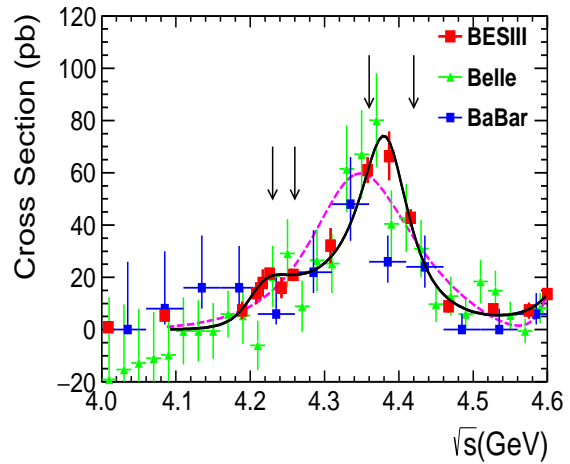}
\setlength{\abovecaptionskip}{-0.2cm}
\setlength{\belowcaptionskip}{-0.6cm}
\caption{Born cross section of $\epem\to\pppm\psip$.
         The dots (red) are the results obtained in this analysis,
         the triangles (green) and squares (blue) are from BELLE and Babar's latest
         updated results, respectively.
         The solid curve is the fit to BESIII results with the coherent sum of three Breit-Wigner functions.
         The dashed curve (pink) is the fit to BESIII results with the coherent sum of two Breit-Wigner functions without the
         $Y(4220)$ hypothesis.
         The arrows mark the locations of four energy points with large luminosities. }
\label{comresult1}
\end{center}
\end{figure}

\section{\boldmath { Fit to the cross section}}
To study the possible resonant structures in $e^{+}e^{-}\to\pi^{+}\pi^{-}\psi(3686)$,
a binned $\chi^{2}$ fit is applied to describe the cross section obtained in this analysis
in a energy region
from 4.085~GeV to 4.600~GeV.
Assuming that three resonances exist,
the PDF can be parameterized as
\begin{equation}\label{eq:2}
\mathcal{A}=f_{1}e^{i\phi_{1}}+f_{2}+f_{3}e^{i\phi_{2}},
\end{equation}
where $f_{1}$ is for the $Y(4220)$, $f_{2}$ is for the $Y(4390)$, $f_{3}$ is for the $Y(4660)$,
$\phi_{1}$ is the phase angle between $Y(4390)$ and $Y(4220)$, and $\phi_{2}$ is the phase angle between $Y(4390)$ and $Y(4660)$.

The amplitude $f_{i}~(i=1,2,3)$ for each resonance is a $P$-wave Breit-Wigner function, defined
as
\begin{equation}\label{eq:3}
  f_{i} = \frac{M_{i}}{M}\frac{\sqrt{12\pi\mathcal{B}_{i}\Gamma^{e^{+}e^{-}}_{i}\Gamma_{i}}}{M^{2}-M^{2}_{i}+iM_{i}\Gamma_{i}}\sqrt{\frac{\Phi(M)}{\Phi(M_{i})}},
\end{equation}
where $M_{i}$ is the mass of the resonance, $\Gamma^{e^{+}e^{-}}_{i}$ is the partial width to $e^{+}e^{-}$, $\Gamma_{i}$ is the total width, $\mathcal{B}_{i}$
is the branching fraction of the resonance decaying to $\pi^{+}\pi^{-}\psi(3686)$, and $\Phi(M_{i})$ is the three-body PHSP factor.
In the fit, $M_{i}$, $\Gamma_{i}$, the product $\mathcal{B}_{i}\Gamma^{e^{+}e^{-}}_{i}$, and the relative phase angle $\phi_{i}$ $(i=1,2)$ are free parameters. The parameters of the $Y(4660)$ are fixed to Belle's latest results~\cite{Y4360Belle}.

The fit result is shown in Fig.~\ref{comresult1} as the solid curve and the resulting parameters of the resonances are summarized in Table~\ref{fitresult}. There are two solutions with equally good fit quality, $\chi^{2}/\text{ndf}=9.98/7$, where $\text{ndf}$ is the number of degrees of freedom. An alternative fit without $Y(4220)$ is performed as shown by the dashed curve in Fig.~\ref{comresult1}, yielding $\chi^{2}/ndf=54.54/11$. The significance of the $Y(4220)$ is determined to be $5.8\sigma$.
\begin{table}[htbp]
\setlength{\belowcaptionskip}{-0.1cm}
  \caption{
  Results of the fit to the $e^{+}e^{-}\to\pi^{+}\pi^{-}\psi(3686)$ cross section. The error is statistical only.
  }\label{fitresult}
  \footnotesize
  \begin{center}
  \begin{tabular}{ c | c  c}
  \hline
  \hline
  ~~~~Parameters ~~~~~~~~~                & ~~~~ Solution I      ~~~~~~~         & ~~~~~~Solution II ~~~~~\\
  \hline
  $M({Y4220})$~~~(MeV/$c^{2}$)                  & \multicolumn{2}{c}{$4209.5\pm7.4$}  \\
  $\Gamma(Y(4220))$~~~(MeV)             & \multicolumn{2}{c}{$80.1\pm24.6$}\\
  $\mathcal{B}\Gamma^{e^{+}e^{-}}(Y(4220))$~~~(eV) & $1.61\pm1.27$   & $1.80\pm1.41$\\
  $M({Y4390})$~~~(MeV/$c^{2}$)                  & \multicolumn{2}{c}{$4383.8\pm4.2$}  \\
  $\Gamma(Y(4390))$~~~(MeV)             & \multicolumn{2}{c}{$84.2\pm12.5$}\\
  $\mathcal{B}\Gamma^{e^{+}e^{-}}(Y(4390))$~~~(eV) & $7.25\pm2.8$    & $10.96\pm3.8$\\
  $\phi_{1}$~~~(rad)                    & $3.3\pm1.0$              & $2.8\pm0.4$\\
  $\phi_{2}$~~~(rad)                    & $0.8\pm0.9$              & $4.7\pm0.1$ \\
  \hline \hline
  \end{tabular}
  \end{center}
  \end{table}

The systematic uncertainties accociated with the resonance parameters include the following.
The uncertainty from the c.m.~energy measurement is studied by taking
the uncertainty of the c.m.~energy $\Delta(\sqrt{s})$=0.8~GeV~\cite{cme} into consideration.
The uncertainty due to the c.m.~energy spread is estimated by convoluting the fit PDF with a Gaussian function with a width of 1.6~MeV, which is the energy spread determined by the Beam Energy Measurement System~\cite{espread}.
The uncertainty from the fit range is investigated by including the first energy point $\sqrt{s}=4.008$~GeV or excluding
the energy point $\sqrt{s}=4.085$~GeV.
The uncertainty from the $Y(4660)$ resonance is studied by varying the parameters of the $Y(4660)$ within its uncertainties.
Table~\ref{sysfitxsec} summarizes the systematic uncertainties for the
mass and total width of the resonances $Y(4220)$ and $Y(4390)$.

\begin{table}[htbp]
\setlength{\belowcaptionskip}{-0.1cm}
  \caption{
  Systematic uncertainty in the measurement of resonance parameters.
  }\label{sysfitxsec}
  \footnotesize
  \begin{center}
  \begin{tabular}{ c | c  c c c }
  \hline
  \hline
  Sources & \multicolumn{2}{c}{$Y(4220)$} & \multicolumn{2}{c}{$Y(4390)$} \\
  & $M$~(MeV/$c^{2})$  & $\Gamma$~(MeV) & $M$~(MeV/$c^{2})$  & $\Gamma$~(MeV) \\
  \hline
  Energy scale  & 0.1  & 0.0  & 0.1 & 0.0 \\
  Energy spread & 1.3  & 0.1  & 0.7 & 1.6 \\
  Fit range     & 0.2  & 1.8  & 0.2 & 0.5  \\
  $Y(4660)$     & 0.6 & 2.3 & 0.3 & 1.2 \\
  \hline
  Total  & 1.4 & 2.9 & 0.8 & 2.1 \\
  \hline \hline
  \end{tabular}
  \end{center}
  \end{table}

A comparison of the masses and widths of the $Y(4220)$ and $Y(4390)$
determined in this work to those of other vector states observed at BESIII
in the mass region between 4.2 and 4.4~GeV/$c^{2}$ are shown in Table~\ref{vectors}.

\begin{table*}[htbp]
\setlength{\belowcaptionskip}{-0.1cm}
  \caption{
  Masses and widths of the vector charmonium states observed from
  different processes at BESIII in
  the mass region between 4.2 and 4.4~GeV/$c^{2}$. The subscript 1 or 2 denotes the
  lower mass state or higher mass state.
  }\label{vectors}
  \footnotesize
  \begin{center}
  \begin{tabular}{ c | c  c c c }
  \hline
  \hline
  Process & ~~~~~$M_{1}$~(MeV/$c^{2}$)~~~~~ & ~~~~~$\Gamma_{1}$~(MeV)~~~~~ & ~~~~~$M_{2}$~(MeV/$c^{2}$)~~~~~ & ~~~~~$\Gamma_{2}$~(MeV)~~~~~\\
  \hline
  $e^{+}e^{-}\to\omega\chi_{c0}$ & $4230\pm8\pm6$ & $38\pm12\pm2$~\cite{omegachicj} \\
  $e^{+}e^{-}\to\pi^{+}\pi^{-}J/\psi$ & $4220.0\pm3.1\pm1.4$ & $44.1\pm4.3\pm2.0$ & $4320.0\pm10.4\pm7.0$ & $101.4^{+25.3}_{-19.7}\pm10.2$\cite{pipijpsi}\\
  $e^{+}e^{-}\to\pi^{+}\pi^{-}h_{c}$ & $4218.4^{+5.5}_{-4.5}\pm0.9$ & $66.0^{+12.3}_{-8.3}\pm0.4$ & $4391.5^{+6.3}_{-6.8}\pm1.0$ & $139.5^{+16.2}_{-20.6}\pm0.6$~\cite{pipihc} \\
  $e^{+}e^{-}\to\pi^{+}D^{0}D^{*-}+c.c$ & $4224.8\pm5.6\pm4.0$ & $72.3\pm9.1\pm0.9$ & $4400.1\pm9.3\pm2.1$ & $181.7\pm16.9\pm7.4$~\cite{piDD} \\
  $e^{+}e^{-}\to\pi^{+}\pi^{-}\psi(3686)$ & $4209.5\pm7.4\pm1.4$ & $80.1\pm24.6\pm2.9$ & $4383.8\pm4.2\pm0.8$ & $84.2\pm12.5\pm2.1$ \\

  \hline \hline
  \end{tabular}
  \end{center}
  \end{table*}

\section{\boldmath { Study of intermediate states}}

Intermediate states in the process $\epem\to\pppm\psip$ are investigated
in the data samples that have large integrated luminosity.
A requirement of $3.68 <M(\pppm\jpsi)(M^{\rm recoil}(\pppm))< 3.70 \;\gevcc$ is applied to extract the $\psip$ signal, and the sideband regions, $3.63 <M(\pppm\jpsi)(M^{\rm recoil}(\pppm))< 3.65\;\gevcc$ or $3.73 <M(\pppm\jpsi)< 3.75 \;\gevcc$, are used to explore the potential
non-$\psip$ backgrounds, where only the left side sideband region is used in mode II since a long tail appears on the right side of the $\psip$ signal due to ISR.
The non-$\psip$ backgrounds are found to be small, and do not produce peaks in the various distributions.

With the above selection criteria, the Dalitz plots of $M^2(\pppm)$ versus $M^2(\pi^\pm\psip)$ and the corresponding one-dimensional projections are shown in Fig.~\ref{intermediate} for data samples at $\sqrt{s} = 4.416, 4.358, 4.258$ and $4.226 \;\gev$, individually, where the plots include the candidates of the two $\psip$ decay modes.
It should be noted that the two higher energy points are from the $Y(4390)$ peak region, while the other two are from the $Y(4220)$ peak region. These four points are marked with vertical arrows in Fig.~\ref{comresult1}.
\begin{figure*}[htbp]
\begin{center}
\begin{overpic}[width=4.8cm,height=4.0cm,angle=0]{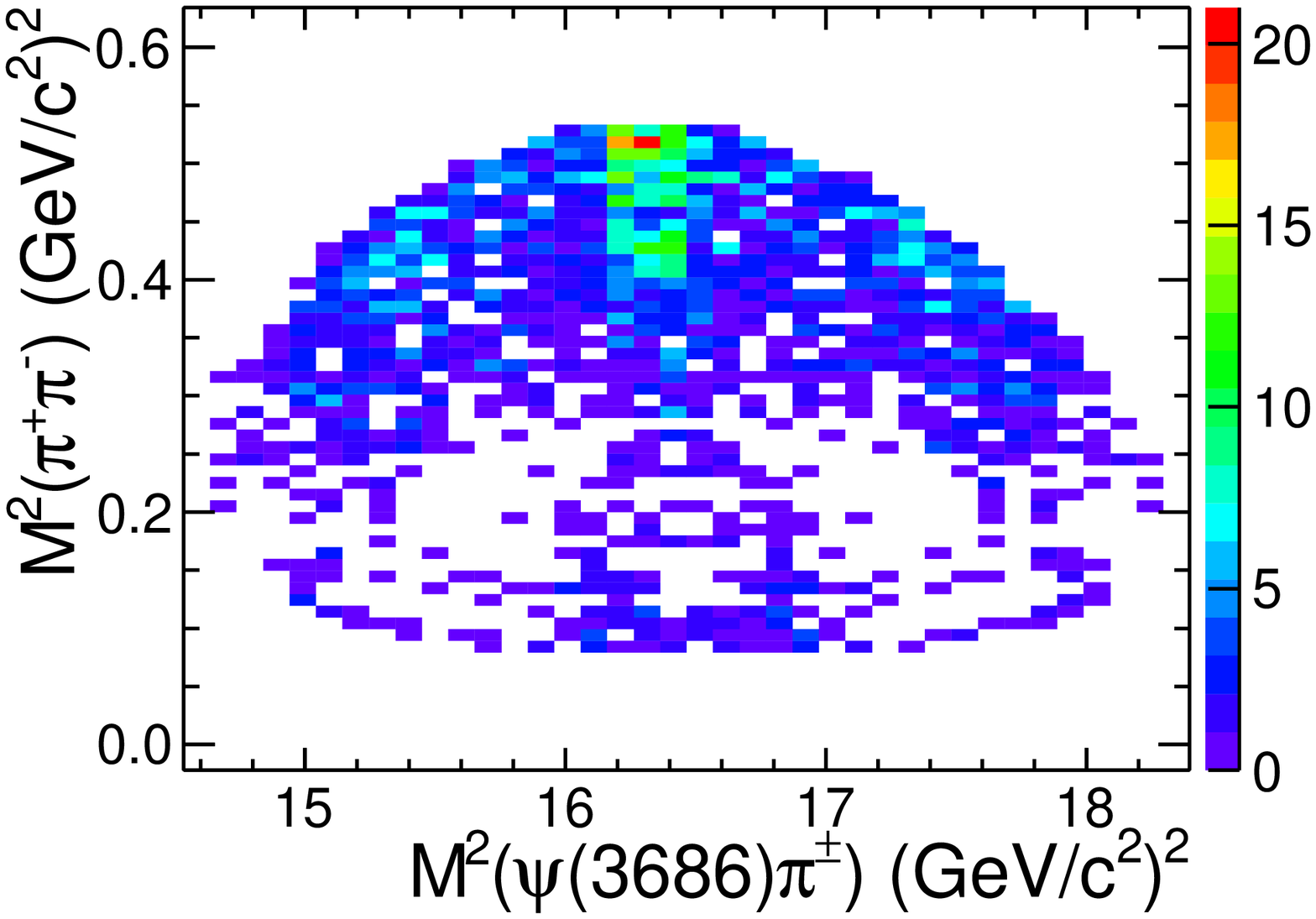}
\put(40,73){\scriptsize{$\sqrt{s}=4.416$~GeV}}
\end{overpic}
\begin{overpic}[width=4.8cm,height=4.0cm,angle=0]{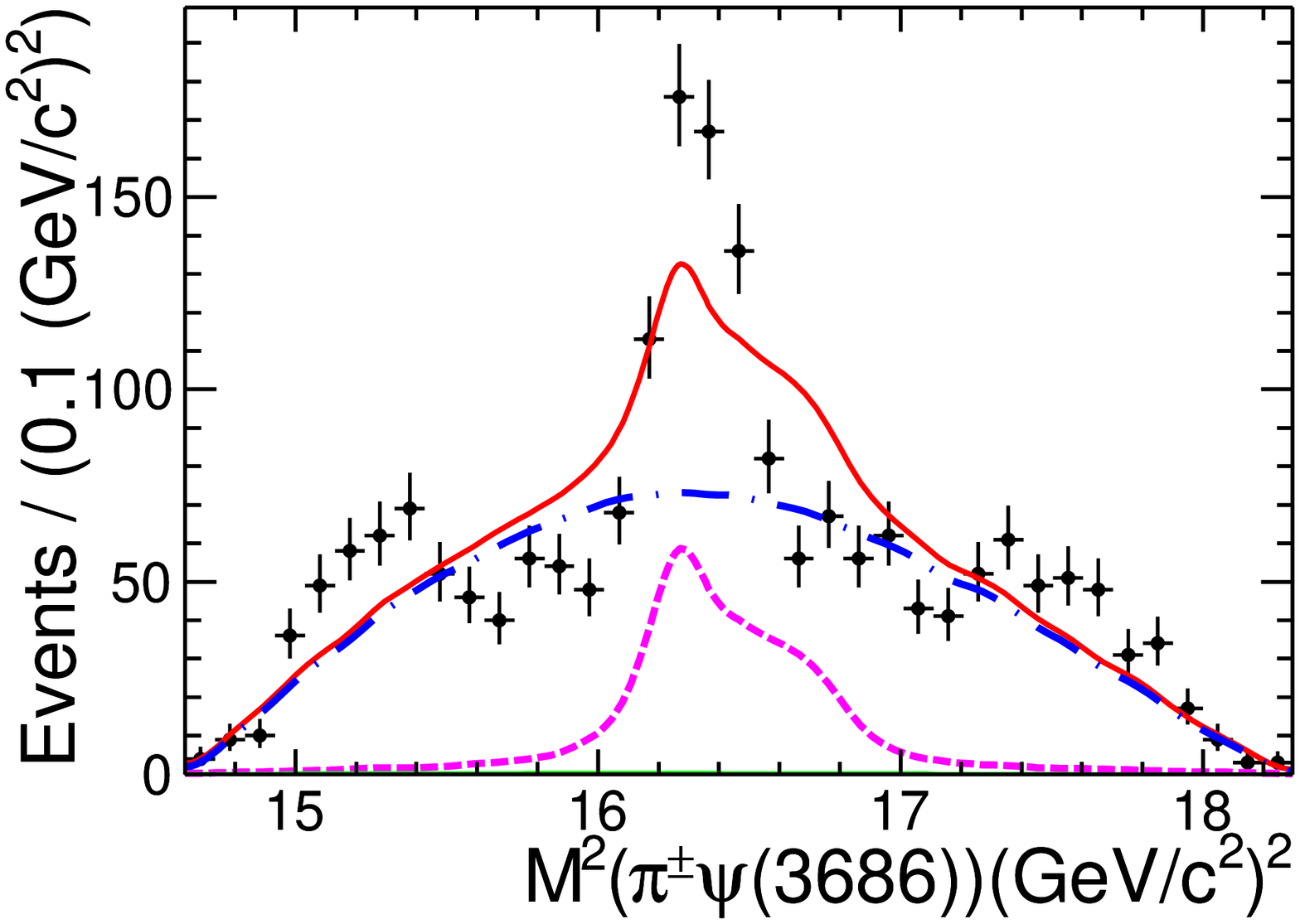}
\end{overpic}
\begin{overpic}[width=4.8cm,height=4.0cm,angle=0]{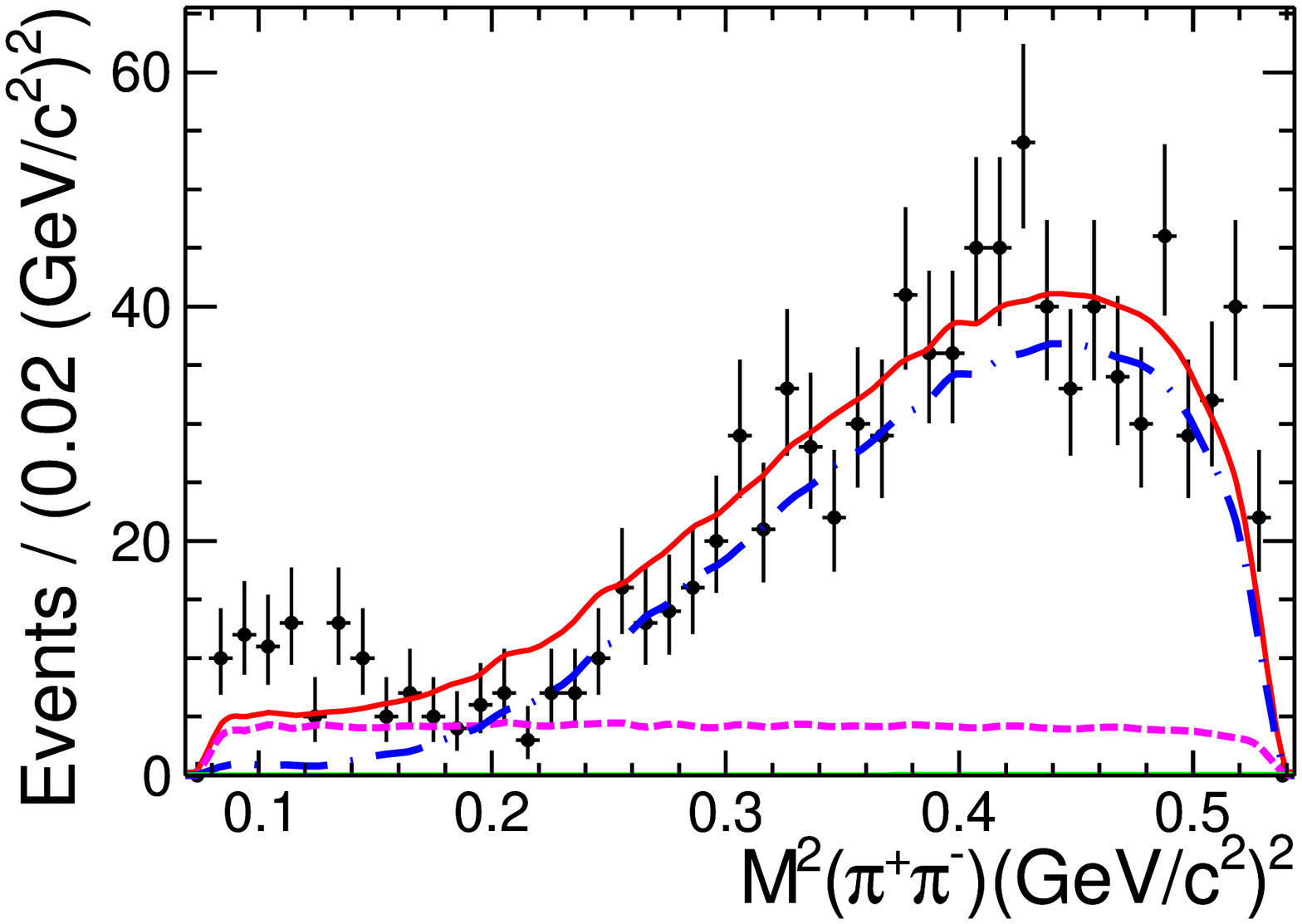}
\end{overpic}\\
\begin{overpic}[width=4.8cm,height=4.0cm,angle=0]{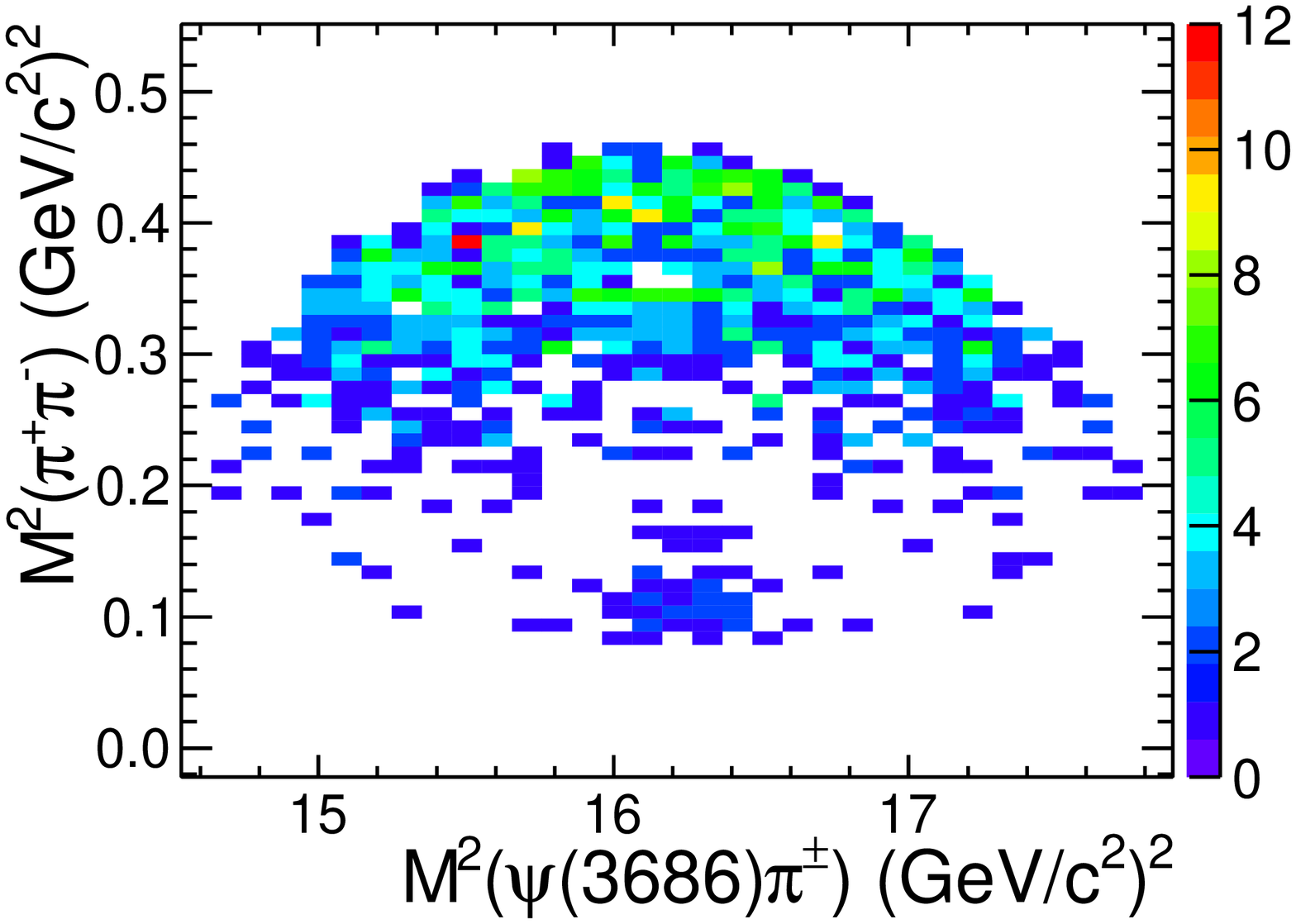}
\put(40,73){\scriptsize{$\sqrt{s}=4.358$~GeV}}
\end{overpic}
\begin{overpic}[width=4.8cm,height=4.0cm,angle=0]{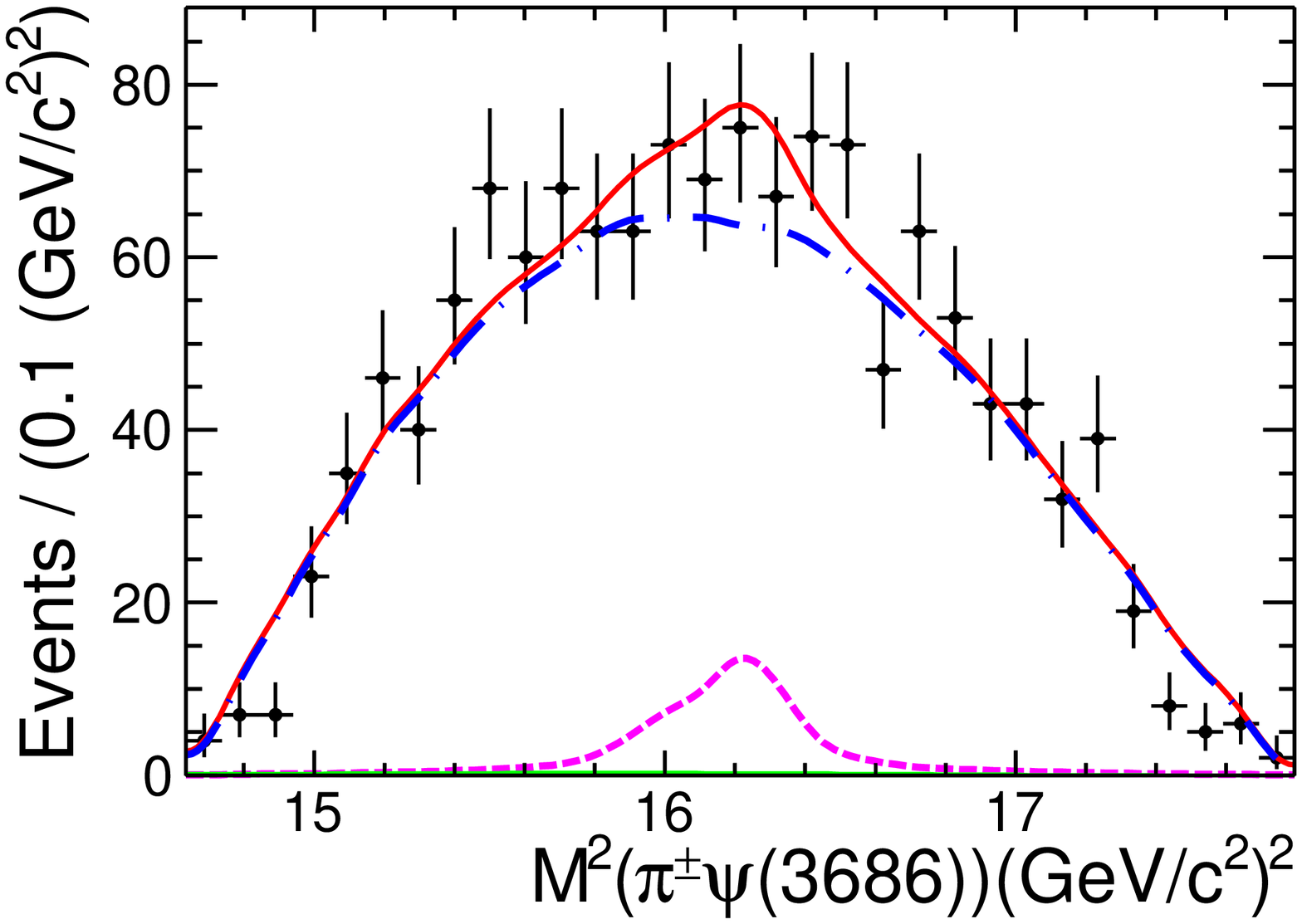}
\end{overpic}
\begin{overpic}[width=4.8cm,height=4.0cm,angle=0]{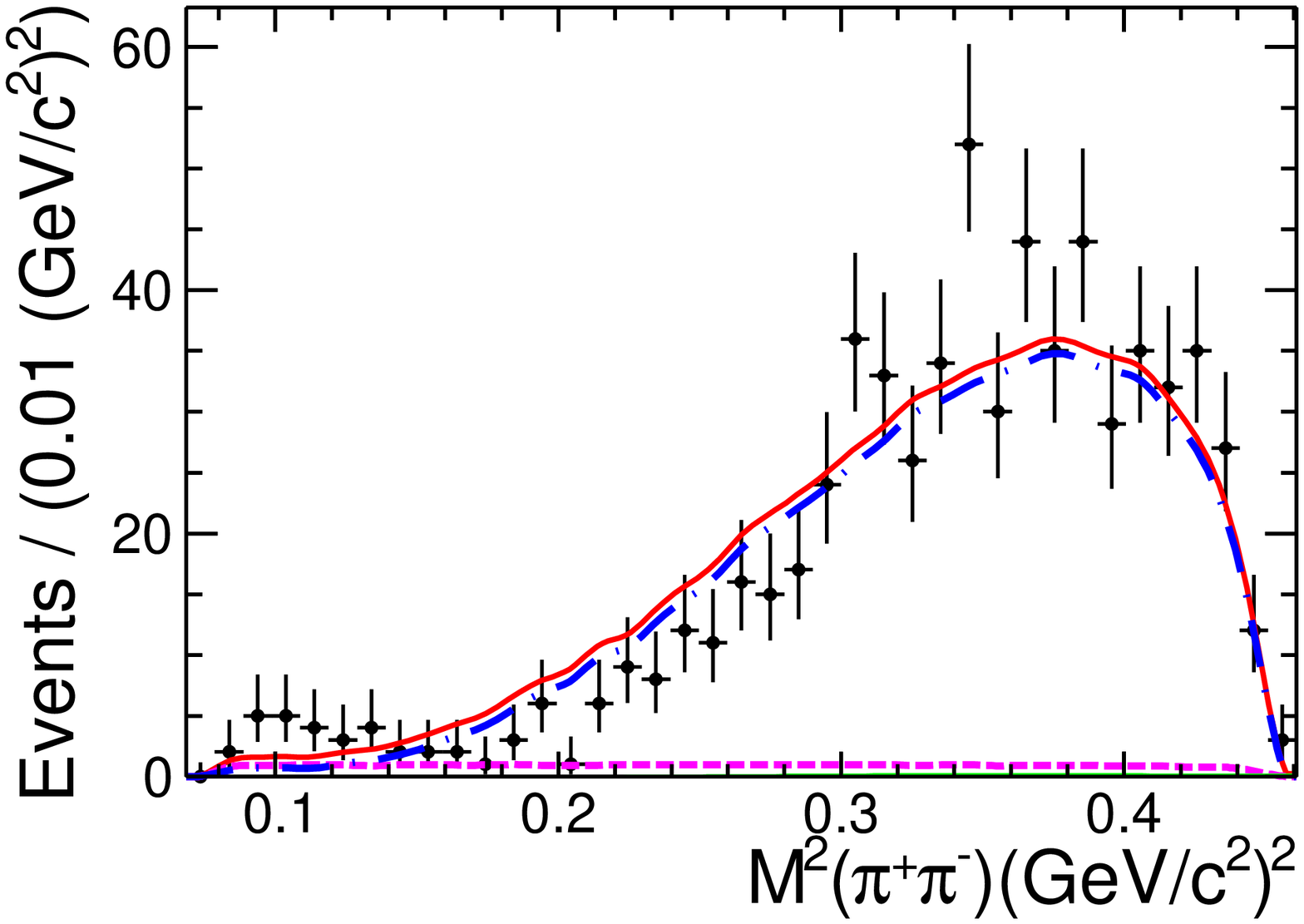}
\end{overpic}\\
\begin{overpic}[width=4.8cm,height=4.0cm,angle=0]{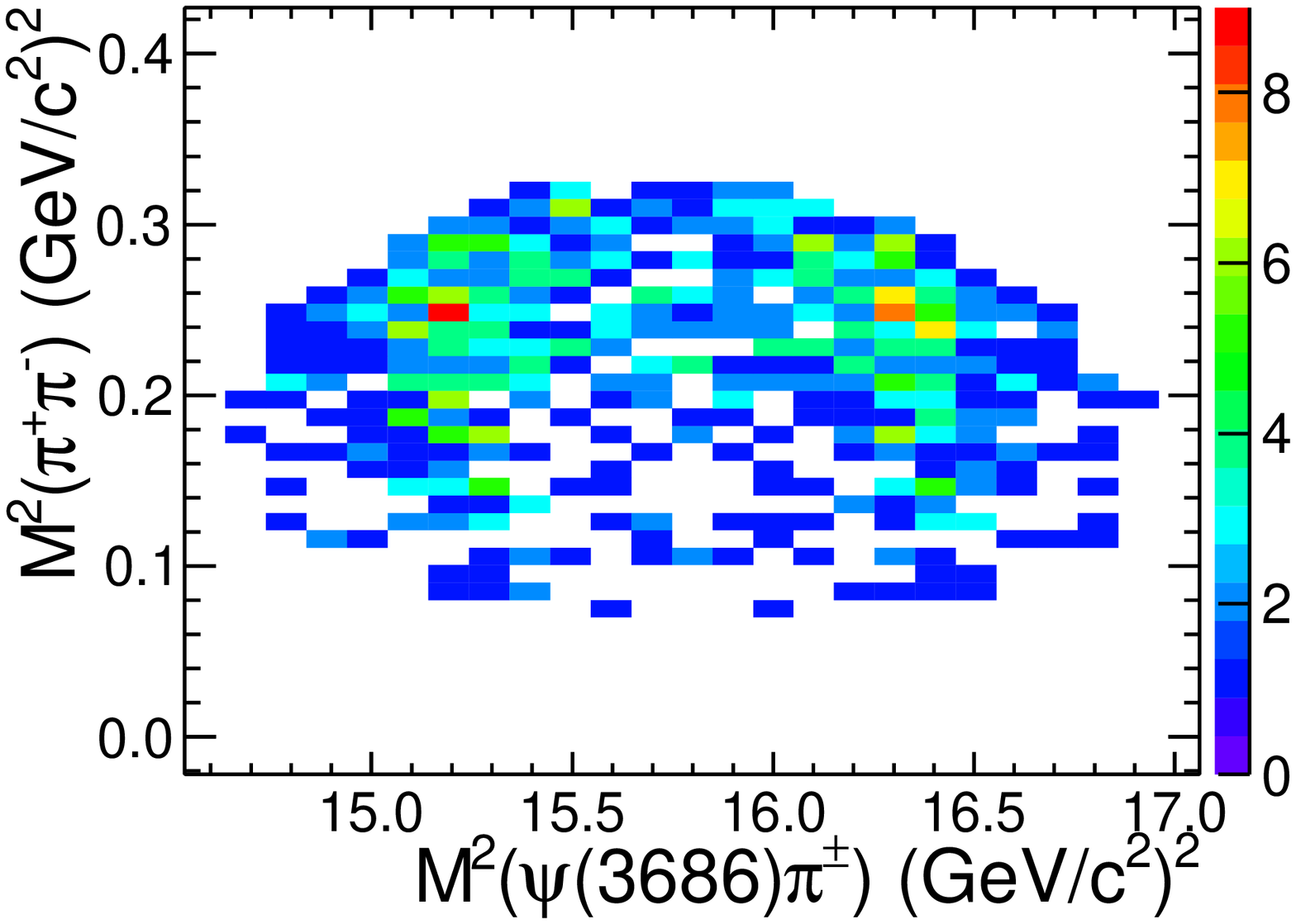}
\put(40,73){\scriptsize{$\sqrt{s}=4.258$~GeV}}
\end{overpic}
\begin{overpic}[width=4.8cm,height=4.0cm,angle=0]{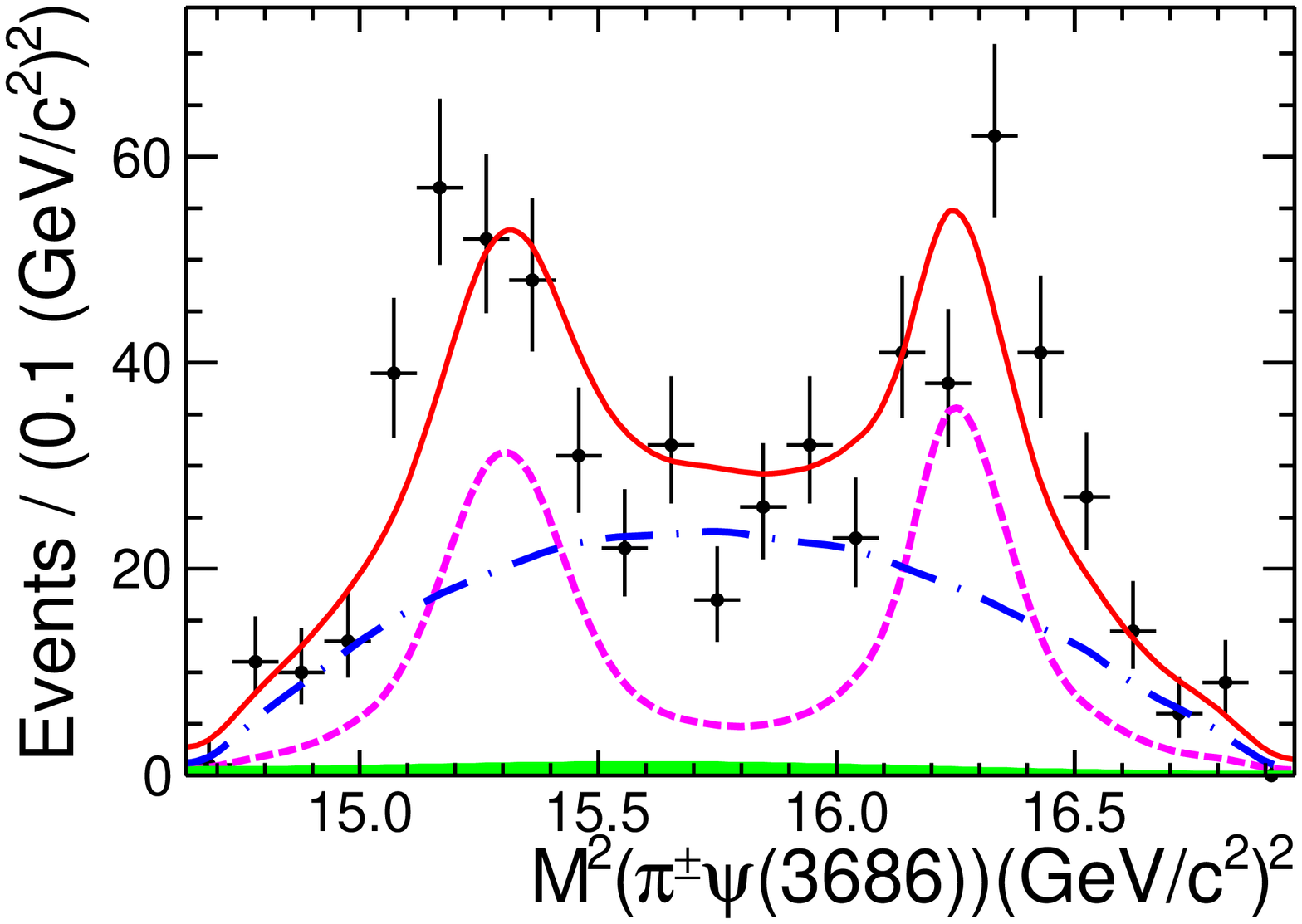}
\end{overpic}
\begin{overpic}[width=4.8cm,height=4.0cm,angle=0]{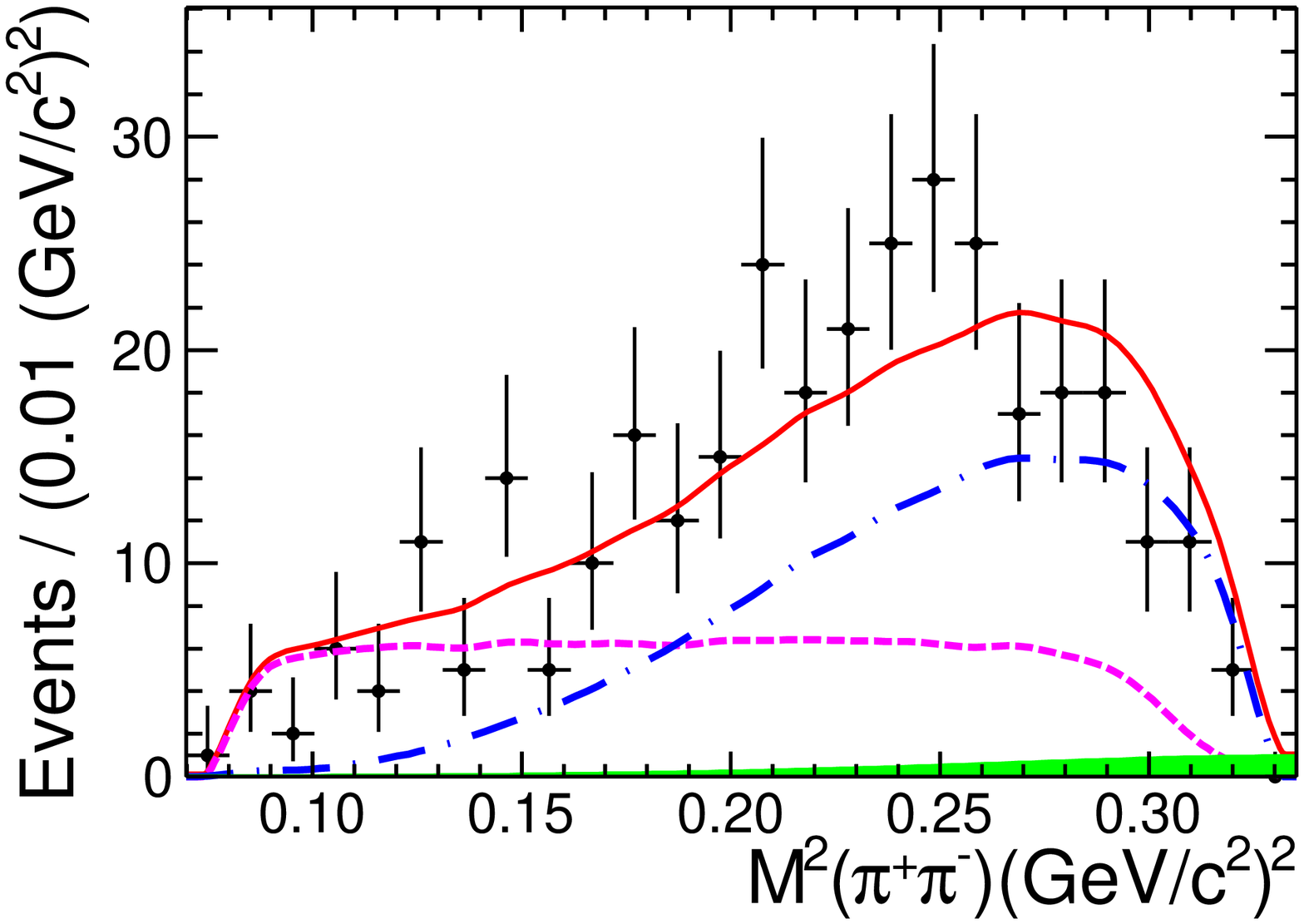}
\end{overpic}\\
\begin{overpic}[width=4.8cm,height=4.0cm,angle=0]{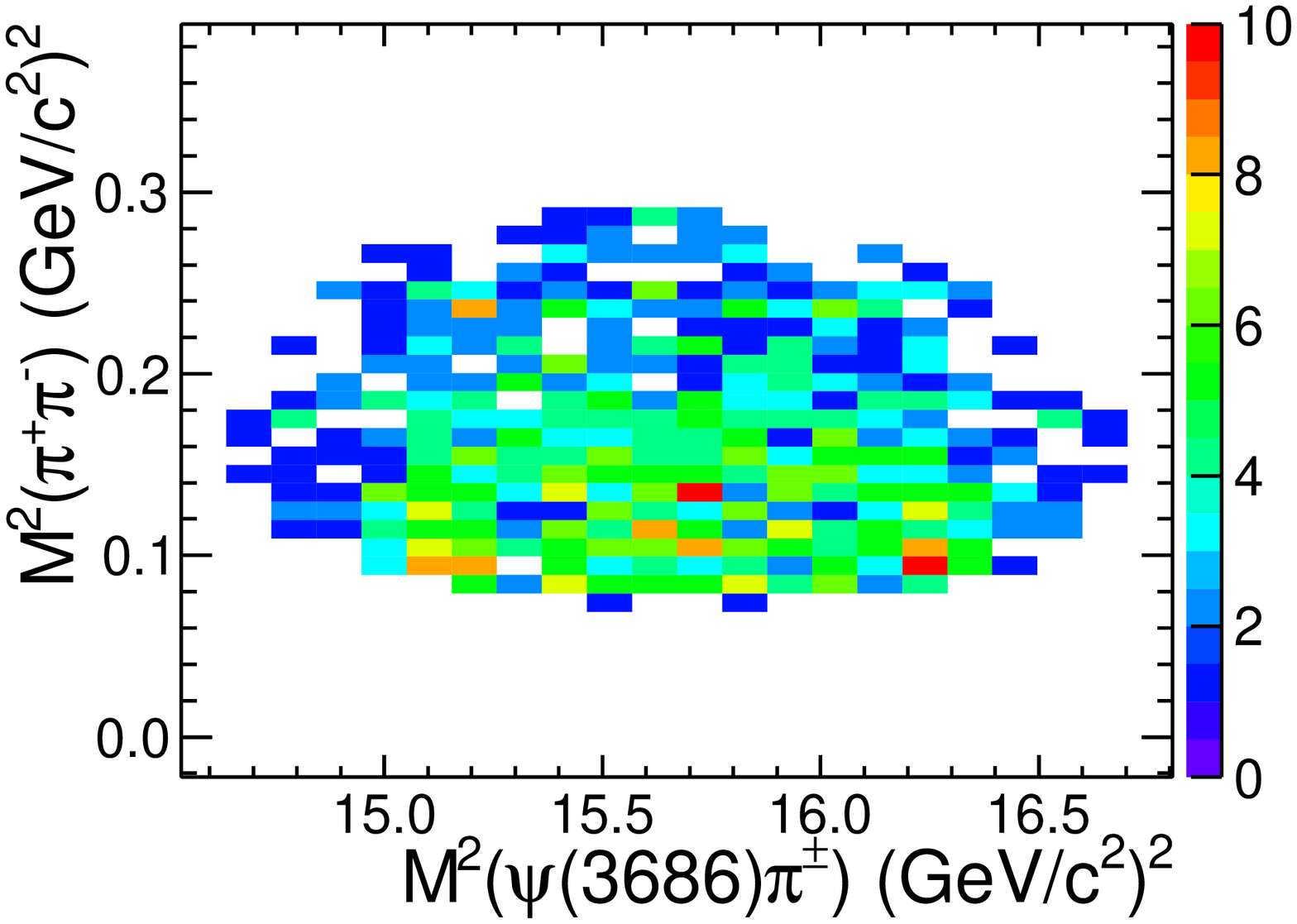}
\put(40,73){\scriptsize{$\sqrt{s}=4.226$~GeV}}
\end{overpic}
\begin{overpic}[width=4.8cm,height=4.0cm,angle=0]{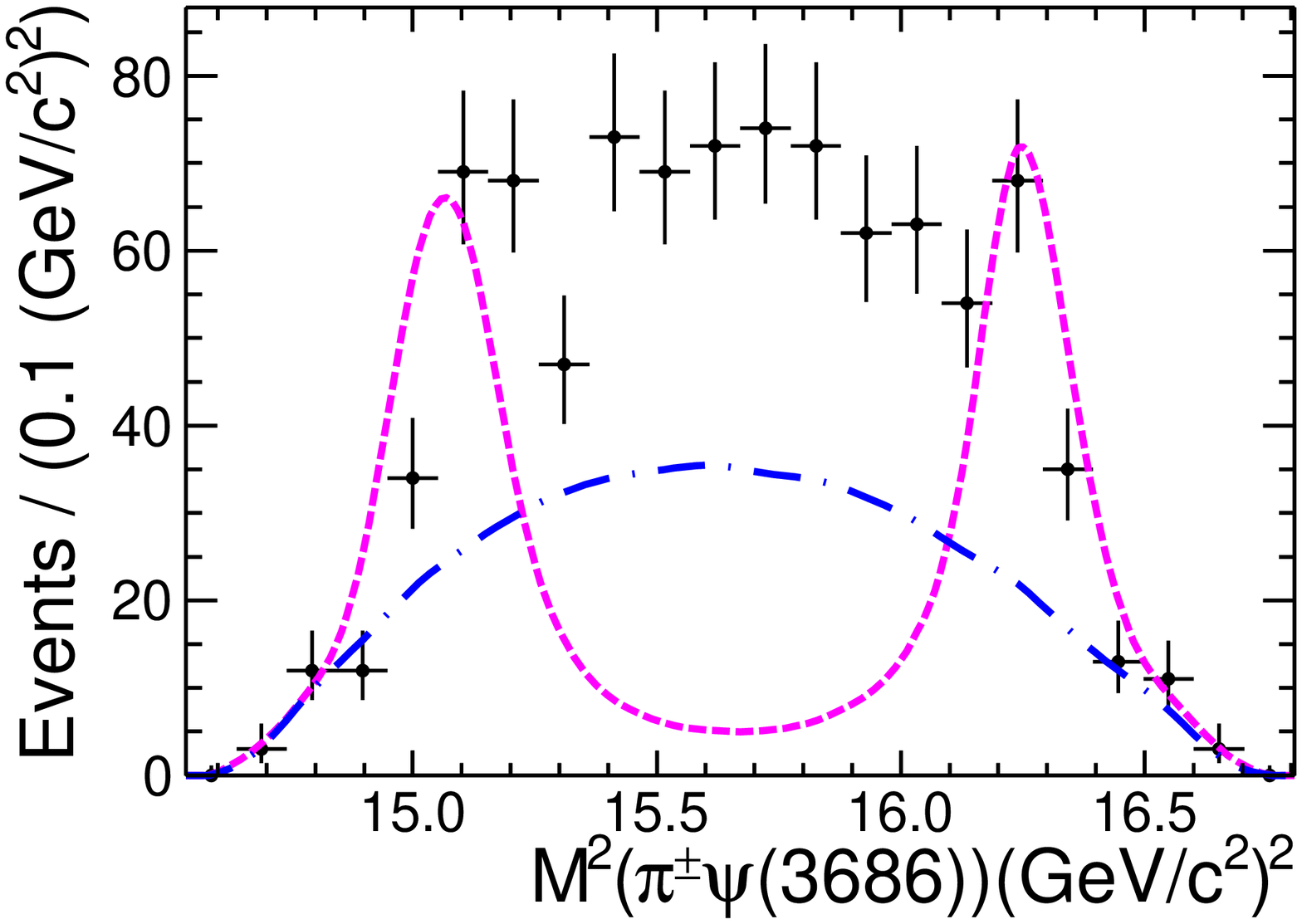}
\end{overpic}
\begin{overpic}[width=4.8cm,height=4.0cm,angle=0]{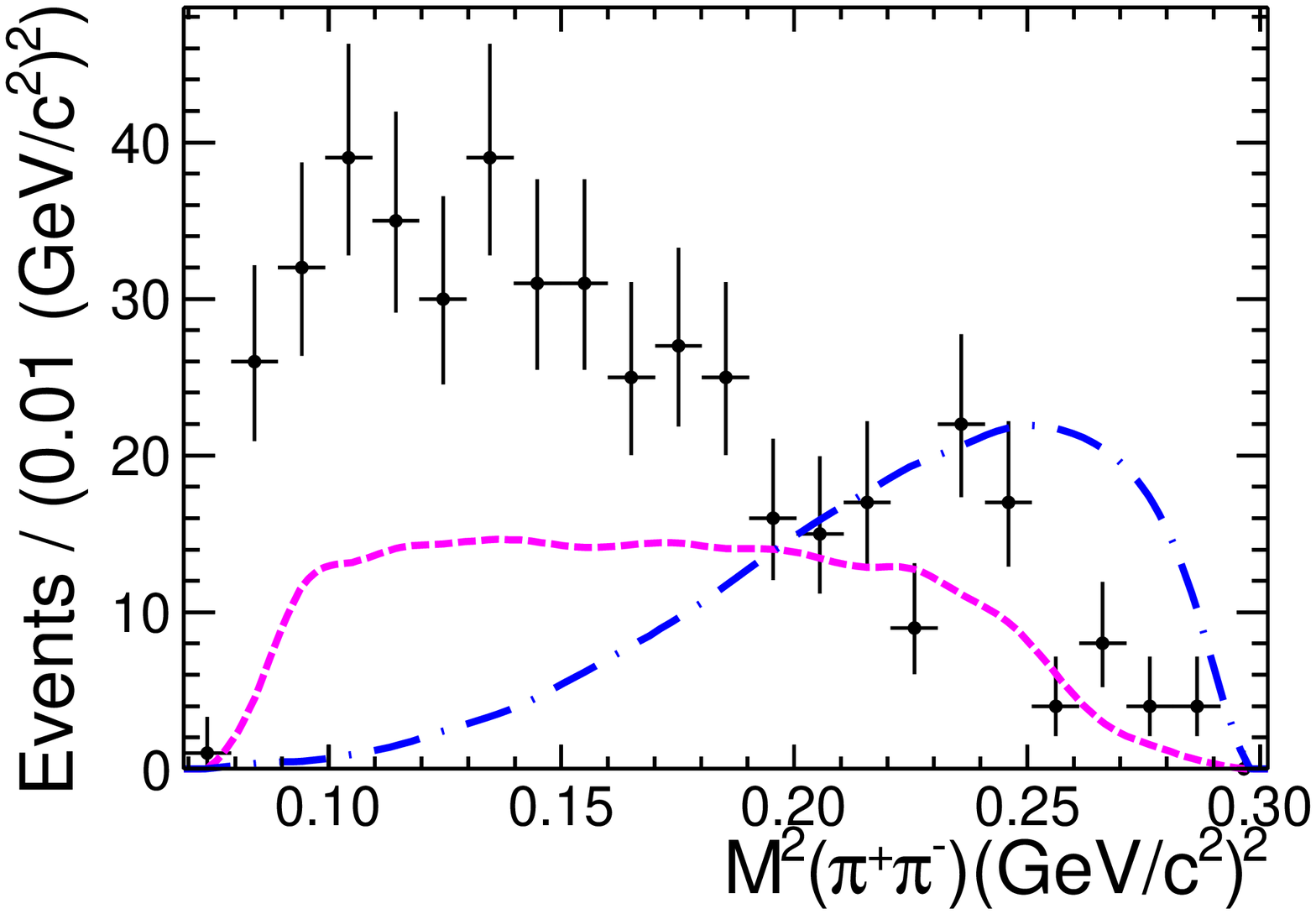}
\end{overpic}\\
\vskip -0.3cm
\setlength{\belowcaptionskip}{-0.5cm}
\parbox[1cm]{16cm} {
\caption{ Dalitz plots of $M^{2}(\pipm\psip)$ versus $M^{2}(\pppm)$,
          distributions of $M^2(\pipm\psip)$ (two entries per event), and $M^{2}(\pppm)$
          for data at $\sqrt{s}= 4.416, 4.358, 4.258$ and $4.226 \;\gev$, with integrated luminosities of 1074, 540, 826 and 1092~pb$^{-1}$, respectively.
          Dots with errors are data.
          For the plots at $\sqrt{s}=4.416, 4.358$ and $4.258\;\gev$,
          the solid curves (red) are projections from the fit;
          the dashed curves (pink) show the shape of the intermediate state;
          the dash-dotted curves (blue) show the shape from the direct process $\epem\to \pppm\psip$ obtained from the \textsc{Jpipi} MC model;
          the shaded histograms (green) show the non-$\psi(3686)$ background estimated with
          the $\psi(3686)$ sideband.
          For plots at $\sqrt{s}=4.226$~GeV, the dashed (pink) and dash-dotted (blue) curves show the shapes from the intermediate state and the direct process $\epem\to\pppm\psip$  (with arbitrary scale).
          In all plots, the two $\psip$ decay modes are combined.
}
\label{intermediate}}
\end{center}
\end{figure*}

For data at $\sqrt{s}= 4.416 \;\gev$, a prominent narrow structure is observed around $4030 \;\mevcc$ in the $M(\pi^\pm\psip)$ spectrum.
The structure is also evident in the corresponding Dalitz plot, but it appears to be more complex when looking at high and low $M(\pppm)$ ranges. In the low $M(\pppm)$ region, there seem to be two separate $\pi^\pm\psip$ structures, presumably corresponding to a physical structure and its kinematic reflection.  But in the high $M(\pppm)$ region, only one broad $\pi^\pm\psip$ structure is seen.
For data at $\sqrt{s}= 4.358 \;\gev$, there is no obvious structure observed in the $M(\pi^\pm\psip)$ spectrum, but a cluster of events appear in the low $M(\pppm)$ region on the corresponding Dalitz plot.
It is worth noting that, at this c.m.~energy, a physical structure with a mass of $4030 \;\mevcc$ in the $M(\pi^\pm\psip)$ spectrum has a reflection at the same mass position.
For data at $\sqrt{s}= 4.258 \;\gev$, two bumps around 3900 and 4030 $\mevcc$ are visible in both the Dalitz plot and in the $M(\pi^\pm\psip)$ spectrum.
For data at $\sqrt{s}= 4.258 \;\gev$, the possible structures with masses of 3900 and 4030 $\mevcc$ in the $M(\pi^\pm\psip)$ spectrum have kinematic reflections at each other's mass positions.
For data at $\sqrt{s}= 4.226 \;\gev$, no structure is clearly seen, which is very different from the behavior at the energy point close by, $\sqrt{s}=4.258~\gev$. A further striking feature for data at $\sqrt{s}= 4.226 \;\gev$  is a very different $M(\pppm)$ distribution from those at the other three energy points.

To characterize the structure observed on the $M(\pi^\pm\psip)$ spectrum for data at $\sqrt{s}= 4.416 \;\gev$, an unbinned maximum likelihood fit is carried out on the Dalitz plot of $M^2(\pip\psip)$ versus $M^2(\pim\psip)$ (denoted as $x$ and $y$ in formula~\ref{eq1}).
Assuming an intermediate state with spin parity $1^{+}$,
the Dalitz plot is parameterized by the coherent sum of the process with an intermediate state and the direct process $\epem\to\pppm\psip$.
The PDF of the intermediate state is described with an $S$-wave Breit-Wigner function without considering interference among the charged conjugate modes,
\begin{equation}\label{eq1}
\frac{p\cdot q/c^2}{(M_{R}^{2}-x)^{2}+M_{R}^{2}\cdot \Gamma^{2}/c^4}+
\frac{p\cdot q/c^2}{(M_{R}^{2}-y)^{2}+M_{R}^{2}\cdot \Gamma^{2}/c^4},
\end{equation}
where $p$ ($q$) is  the $\psip$ (intermediate state) momentum in the $\pipm\psip$ (initial $\epem$) rest frame, and $M_R$ and $\Gamma$ are the mass and width of the intermediate state.
The 2-dimensional mass resolution and the detection efficiency,
determined from MC simulations, are incorporated in the PDF for the intermediate states in the fit.
The PDF of the direct process $\epem\to\pppm\psip$ is taken from a MC-simulated shape using the \textsc{Jpipi} model, and that of the non-$\psip$ background is described with the distribution of events in the $\psip$ sideband region.
A simultaneous fit constraining the mass and width of the intermediate state is carried out by maximize the product of the likelihood values of the two $\psip$ decay modes.
The fit process is validated using MC samples.
The distributions of mass resolution and detection efficiency are provided in the appendix.

The fit yields a mass of $M = 4032.1\pm2.4 \;\mevcc$ and a width of $\Gamma$ = $26.1\pm5.3 \;\mev$ for the intermediate state with a significance of 9.2$\sigma$, evaluated by comparing the likelihood values with or without the intermediate states included.
The fit projections on $M^{2}(\pi^{\pm}\psip)$ and $M^{2}(\pppm)$ for data at $\sqrt{s}=4.416 \;\gev$ are shown in Fig.~\ref{intermediate}.
It can be seen that the overall fit curve does not match the peaking structure on the $M(\pi^\pm\psip)$ spectrum in data, and the corresponding confidence level (C.L.) of the fit is only 8\%, as estimated by toy-MC tests.
Alternative fits with different assumptions of the spin parity of the intermediate state, including the interference among the charge conjugated modes, and including the contribution of $\zcbs$ are explored.
In these fits the parameters of the intermediate state are close to
the norminal fit result and the fit qualities are not improved significantly.
As shown in the Dalitz plot, the behavior of the structure is very different between the high $M(\pppm)$ region and the low $M(\pppm)$ region. A similar fit to data with the additional requirement $M^2(\pppm)> 0.3 \;(\gevcc)^2$ is performed, which yields a mass of $M=4030.3\pm0.1 \;\mevcc$ and a width of $\Gamma = 5.1\pm0.2 \;\mev$.
The corresponding projection of the fit and data on the $M^2(\pi^\pm\psip)$ distribution is shown in Fig.~\ref{intermediate2}, and the fit C.L. is  50\%.
\begin{figure}[htbp]
\begin{center}
\begin{overpic}[width=7.6cm,height=6.5cm,angle=0]{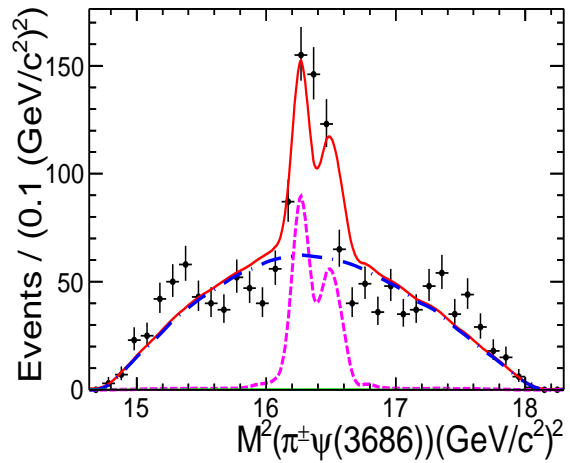}
\end{overpic}
\setlength{\belowcaptionskip}{-0.6cm}
\caption{ Projection of $M^2(\pipm\psip)$ at $\sqrt{s}=4.416$~GeV with a requirement of
$M^2(\pppm)>0.3 \;(\gevcc)^2$.
}
\label{intermediate2}
\end{center}
\end{figure}

Similar fits are carried out to data at $\sqrt{s}= 4.358$ and $4.258 \;\gev$, where the parameters of the intermediate state are fixed to those from the fit to data at $\sqrt{s}= 4.416 \;\gev$.
The projections on $M^{2}(\pi^{\pm}\psip)$ and $M^{2}(\pppm)$ are also shown in
Fig.~\ref{intermediate}.
The statistical significance of the intermediate state is $3.6\sigma$ and $9.6\sigma$ for data at $\sqrt{s}= 4.358$ and $4.258 \;\gev$, respectively.
For data at $\sqrt{s}=4.358$~GeV, as shown in the Dalitz plot, a cluster of events appears in the $M^{2}(\pi^{\pm}\psi(3686))$ spectrum
at low $M^{2}(\pi^{+}\pi^{-})$, which also shows a hint of an intermediate structure.
As mentioned previously,
for data at $\sqrt{s}= 4.258 \;\gev$, the structures with masses of 3900 and 4030 $\mevcc$ on the $M(\pi^\pm\psip)$ spectrum are kinematic reflections of each other, so the fit results are strongly dependent on whether the $\zcbs$ is included in the fit or not.
For data at $\sqrt{s} = 4.226 \;\gev$, an intermediate state with a mass of 4030~MeV and its reflection are very close to the kinematic boundary of the three-body decay, so no obvious peak is observed in the $M(\pipm\psip)$ spectrum. The anomalous distribution on the $M(\pppm)$ spectrum is, however, discussed in Ref.~\cite{zoubs}.
For the other energy points with high statistics, such as $\sqrt{s}=4.387$ and 4.600~GeV, the Dalitz plots and the distribution of $M^{2}(\pi^{\pm}\psi(3686))$
and $M^{2}(\pi^{+}\pi^{-})$ are shown in Fig.~\ref{intermediate3}.

\begin{figure*}[htbp]
\begin{center}
\begin{overpic}[width=4.8cm,height=4.0cm,angle=0]{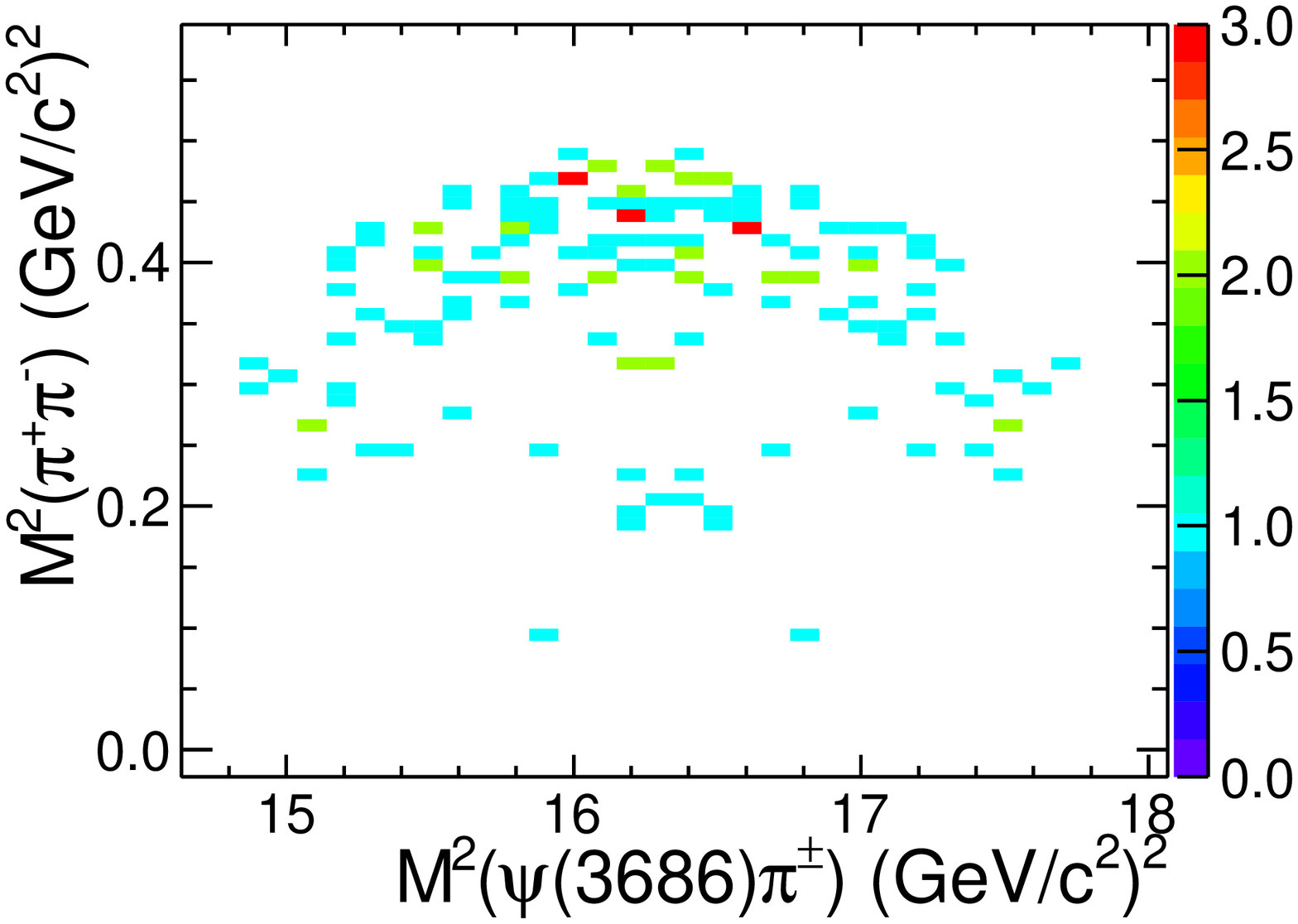}
\put(40,73){\scriptsize{$\sqrt{s}=4.387$~GeV}}
\end{overpic}
\begin{overpic}[width=4.8cm,height=4.0cm,angle=0]{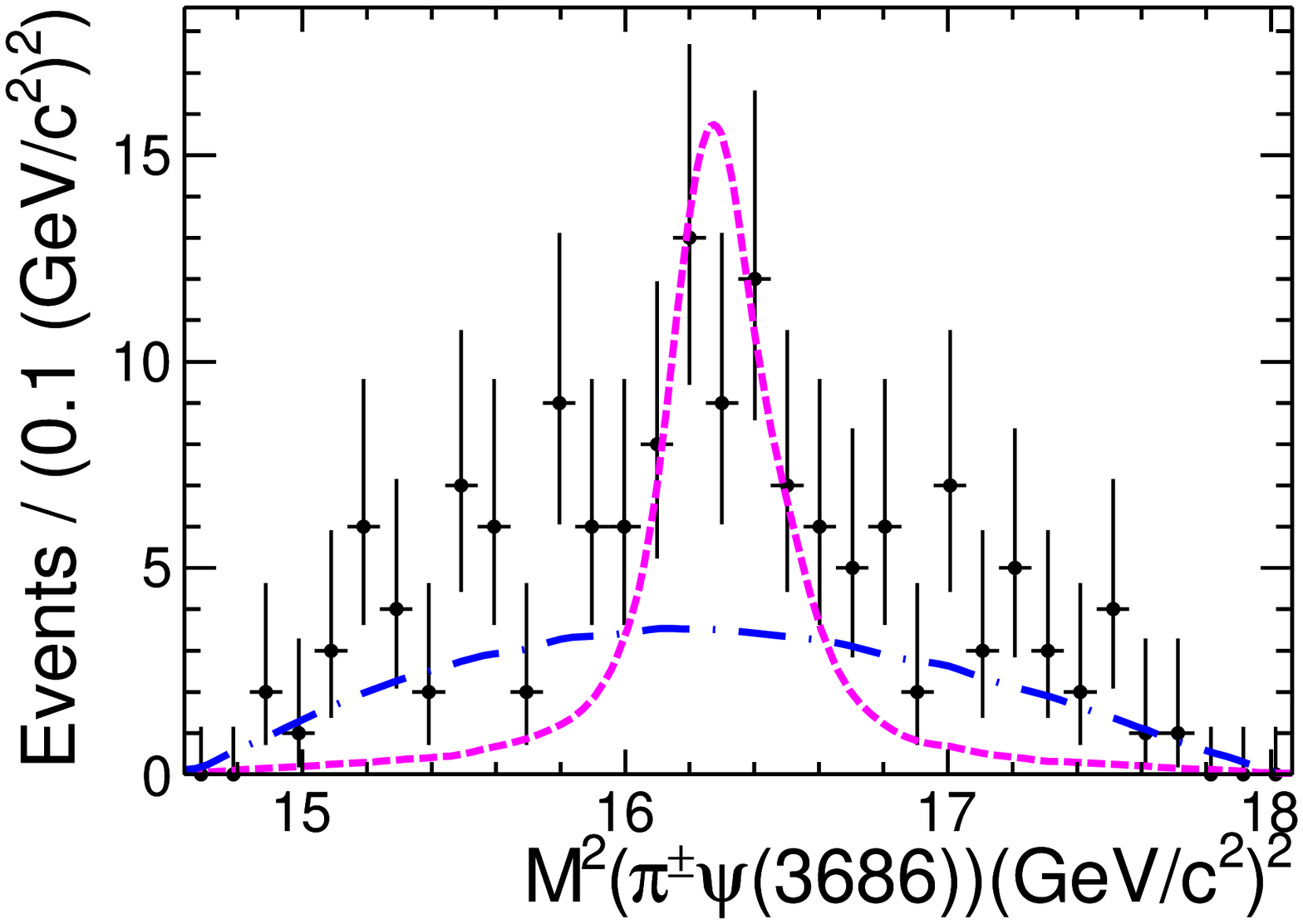}
\end{overpic}
\begin{overpic}[width=4.8cm,height=4.0cm,angle=0]{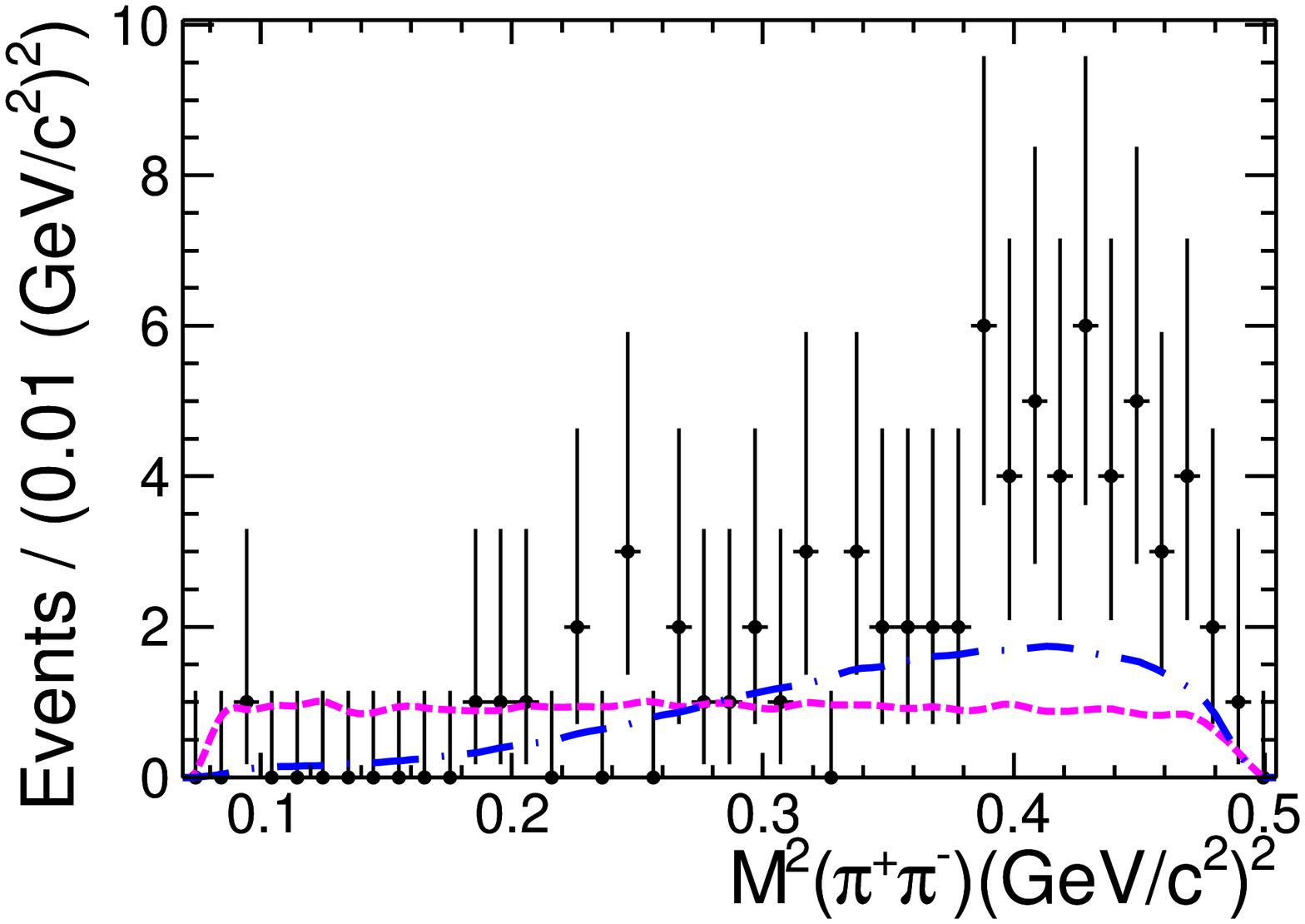}
\end{overpic}\\
\begin{overpic}[width=4.8cm,height=4.0cm,angle=0]{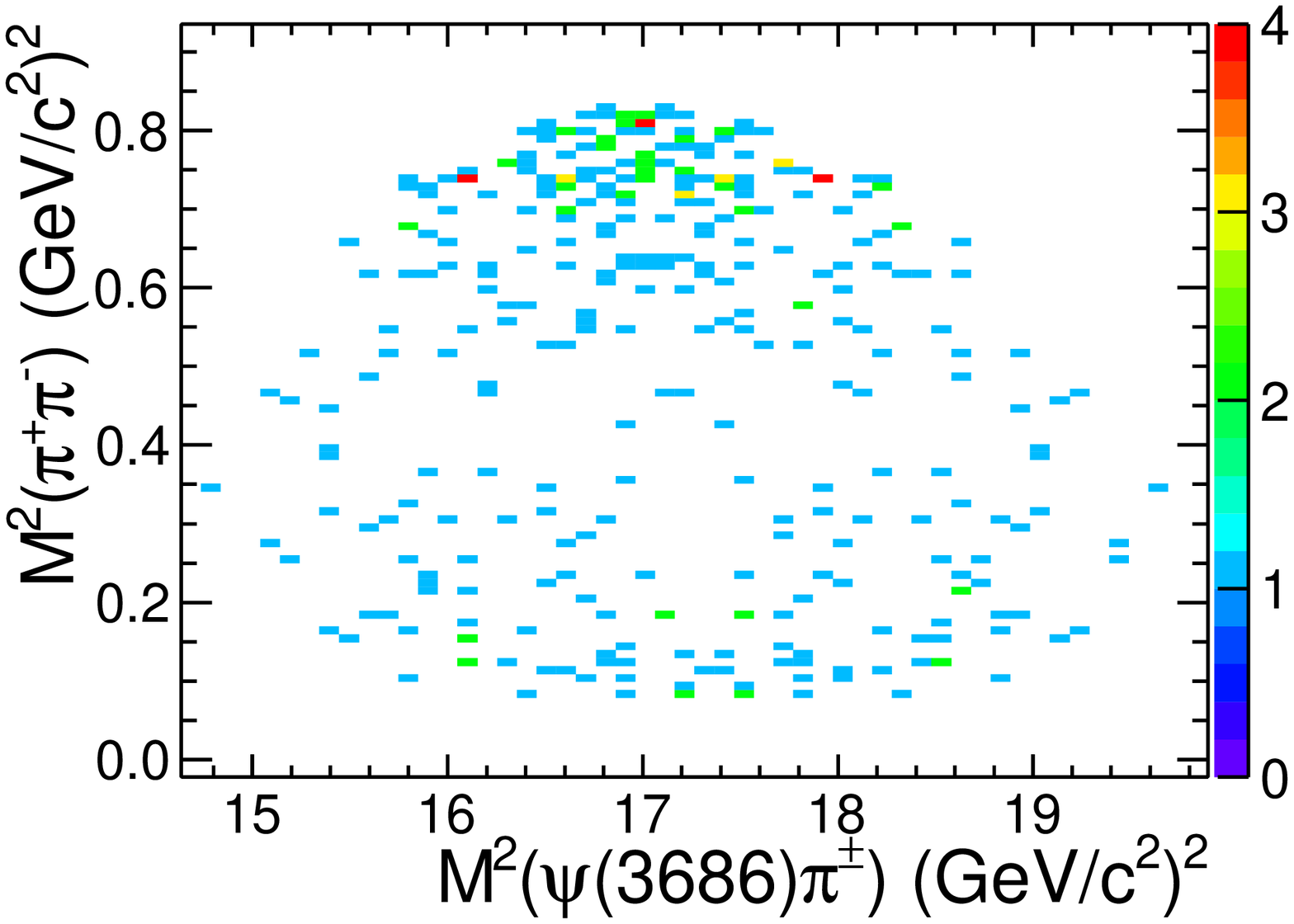}
\put(40,73){\scriptsize{$\sqrt{s}=4.600$~GeV}}
\end{overpic}
\begin{overpic}[width=4.8cm,height=4.0cm,angle=0]{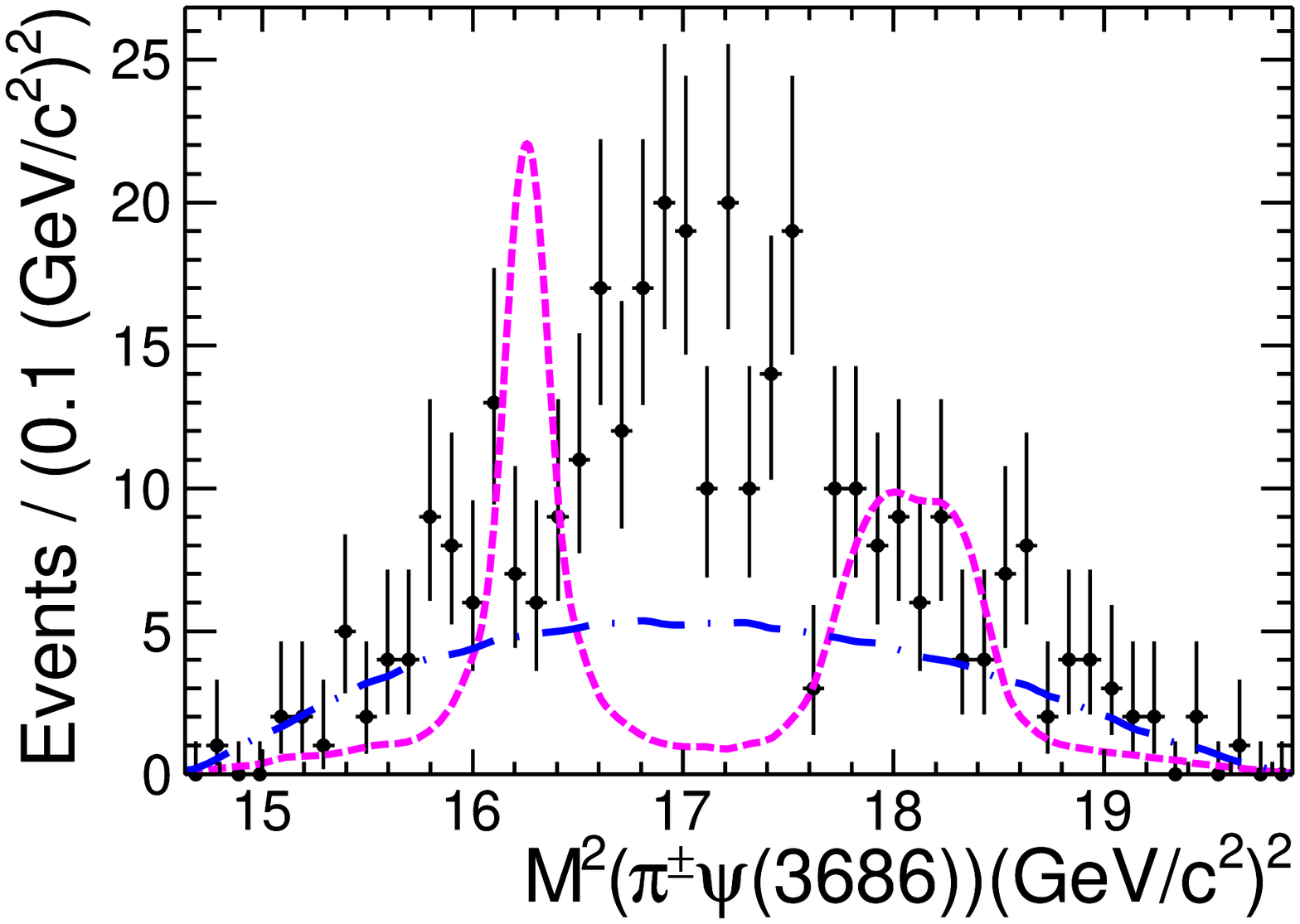}
\end{overpic}
\begin{overpic}[width=4.8cm,height=4.0cm,angle=0]{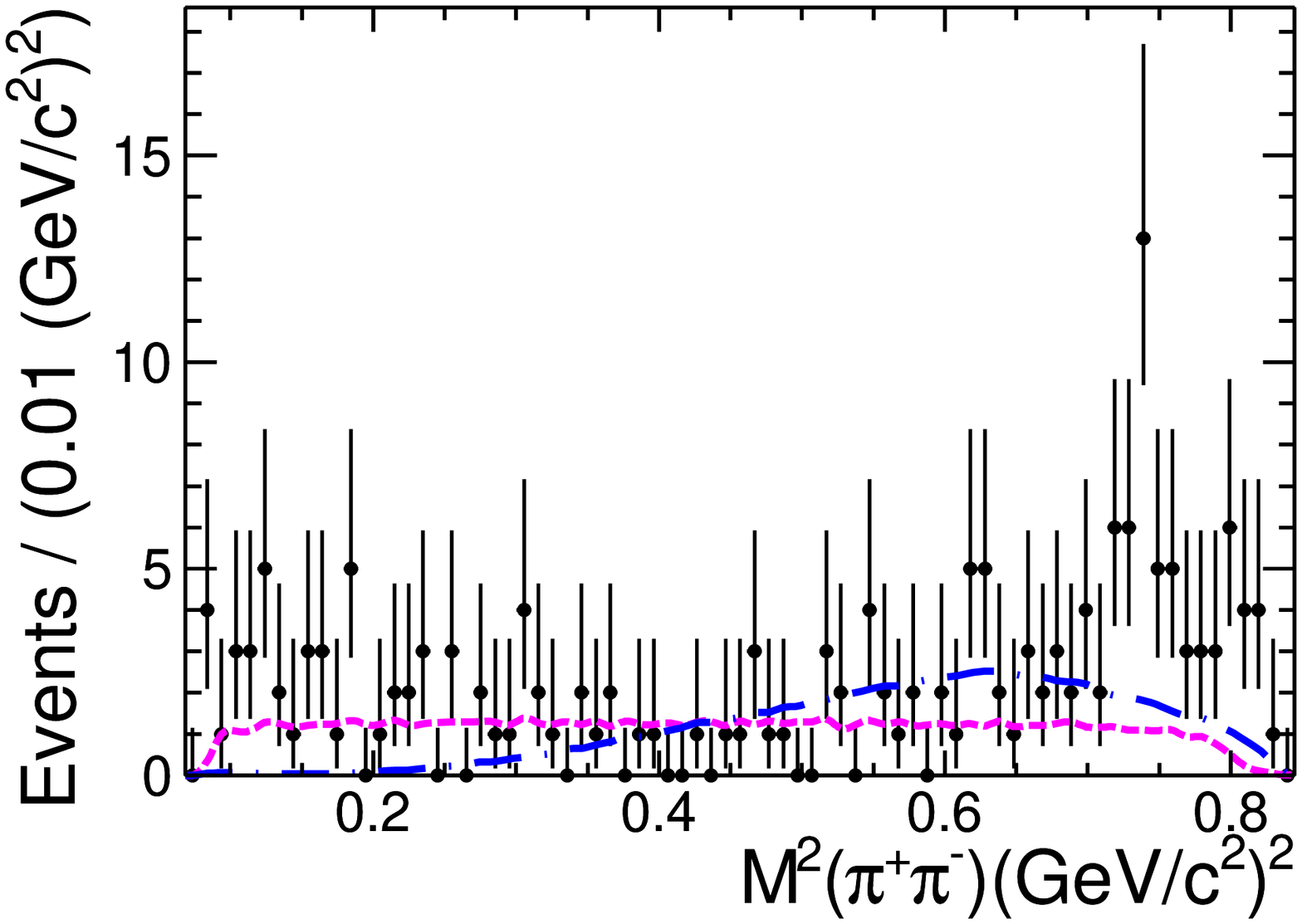}
\end{overpic}\\
\setlength{\abovecaptionskip}{-0.2cm}
\setlength{\belowcaptionskip}{-0.6cm}
\caption{
Dalitz plots of $M^{2}(\pipm\psip)$ versus $M^{2}(\pppm)$,
          distributions of $M^2(\pipm\psip)$ (two entries per event), and $M^{2}(\pppm)$
          for data at $\sqrt{s}= 4.38$ and 4.600$\;\gev$,
           Dots with errors are data, the dashed (pink) and dash-dotted (blue) curves show the shapes from intermediate state and direct process $\epem\to\pppm\psip$  (with arbitrary scale).
          }
\label{intermediate3}
\end{center}
\end{figure*}

\section{SUMMARY}

In summary, based on 5.1~$\ifb$ of $\epem$ collision data with c.m.~energies between 4.008 and 4.600 $\gev$, collected with the BESIII detector, the Born cross sections of $\epem\to\pppm\psip$ are measured.
The measured cross sections are in good agreement with those from Babar and Belle~\cite{Y4360BaBar,Y4360Belle}, but with significantly improved precision.
The cross section is fit with three coherent Breit-Wigner functions.
The parameters of two resonances, the $Y(4220)$ and $Y(4390)$, are determined from the fit to be
$M=4209.5\pm7.4\pm1.4$~MeV/$c^{2}$ and $\Gamma=80.1\pm24.6\pm2.9$~MeV for the $Y(4220)$,
and $M=4383.8\pm4.2\pm0.8$~MeV/c$^{2}$ and $\Gamma=84.2\pm12.5\pm2.1$~MeV for the $Y(4390)$,
where the first errors are statistical and the second systematic.
The resonance $Y(4220)$ is observed in the process $e^{+}e^{-}\to\pi^{+}\pi^{-}\psi(3686)$
for the first time with a significance of $5.8\sigma$.
Both of the resonances $Y(4220)$ and $Y(4390)$ are
consistent with the vector resonances
observed in $e^{+}e^{-}\to\pi^{+}\pi^{-}h_{c}$~\cite{pipihc}.
The $Y(4220)$ state is possibly the same state as observed
in $e^{+}e^{-}\to\pi^{+}\pi^{-}J/\psi$~\cite{pipijpsi} and $e^{+}e^{-}\to \omega\chi_{c0}$~\cite{omegachicj},
and an attempt was made with a combined fit to
the $e^{+}e^{-}\to \omega\chi_{c0}$, $\pi^{+}\pi^{-}J/\psi$, 
$\pi^{+}\pi^{-}h_{c}$,
and $\pi^{+}D^{0}D^{*-}+c.c$ cross sections measured by BESIII to determine 
the resonant parameters of the $Y(4220)$~\cite{combinefit}.
Peaking at a similar mass was also observed in the process
$e^{+}e^{-}\to\eta J/\psi$~\cite{etajpsi} at BESIII.

A charged charmonium-like structure is observed in the $M(\pi^\pm\psip)$ spectrum for data at $\sqrt{s} = 4.416 \;\gev$.
A fit to the structure, assuming the existence of a spin-parity $1^+$ charmonium-like state, yields a mass $M=4032.1\pm2.4 \;\mevcc$,
where the errors are statistical only.
However, the fit cannot describe data well.
The width of the intermediate state varies over a wide range for different kinematic regions of data at $\sqrt{s}=4.416$~GeV.
Similar fits are carried out to data at $\sqrt{s}= 4.258$ and $4.358
\;\gev$, where there is also evidence for the new structure at 4030 $\mevcc$ in the $M(\pi^\pm\psip)$ spectra.
For data at $\sqrt{s}= 4.226 \;\gev$, no obvious structure in the $M(\pi^\pm\psip)$ spectrum is observed, but the $M(\pppm)$ spectrum shows an anomalous distribution.
The measured mass of the intermediate state deviates from that of the
structure observed by Belle with their full data sample~\cite{Y4360Belle} by over $3\sigma$.
Larger data sets, as well as additional theoretical input, are necessary for further understanding of the intermediate structures in $\epem\to\pppm\psip$.

\section{ACKNOWLEDGMENTS}
The BESIII collaboration thanks the staff of BEPCII, the IHEP computing center and the supercomputing center of USTC for their strong support. This work is supported in part by National Key Basic Research Program of China under Contract No. 2015CB856700; National Natural Science Foundation of China (NSFC) under Contracts Nos. 11235011, 11322544, 11335008, 11375170, 11275189, 11425524, 11475164, 11475169, 11625523, 11605196, 11605198, 11635010; the Chinese Academy of Sciences (CAS) Large-Scale Scientific Facility Program; the CAS Center for Excellence in Particle Physics (CCEPP); Joint Large-Scale Scientific Facility Funds of the NSFC and CAS under Contracts Nos. U1332201, U1532257, U1532258, U1532102; CAS under Contracts Nos. KJCX2-YW-N29, KJCX2-YW-N45, QYZDJ-SSW-SLH003; 100 Talents Program of CAS; National 1000 Talents Program of China; INPAC and Shanghai Key Laboratory for Particle Physics and Cosmology; German Research Foundation DFG under Contracts Nos. Collaborative Research Center CRC 1044, FOR 2359; Istituto Nazionale di Fisica Nucleare, Italy; Joint Large-Scale Scientific Facility Funds of the NSFC and CAS; Koninklijke Nederlandse Akademie van Wetenschappen (KNAW) under Contract No. 530-4CDP03; Ministry of Development of Turkey under Contract No. DPT2006K-120470; National Natural Science Foundation of China (NSFC); National Science and Technology fund; The Swedish Resarch Council; U. S. Department of Energy under Contracts Nos. DE-FG02-05ER41374, DE-SC-0010118, DE-SC-0010504, DE-SC-0012069; University of Groningen (RuG) and the Helmholtzzentrum fuer Schwerionenforschung GmbH (GSI), Darmstadt; WCU Program of National Research Foundation of Korea under Contract No. R32-2008-000-10155-0.\\

\appendix
\section{ Distribution of $\epsilon(x,y)$ and $\sigma(x,y)$ }
In Eq.~\ref{eq2}, the fit procedure can be carried out by the
Breit-Wigner function for $M^2(\pip\psip)$ and $M^2(\pim\psip)$
(denoted as $x$ and $y$) convoluted with a
resolution function $\sigma(x)$, $\sigma(y)$, respectively,
and then each of them multiplies by a flat polynomial on $y$ and $x$
to form a 2-dimensional function.
The signal shape can then be obtained by multiplying a 2-dimensional resolution function $\epsilon(x,y)$
after adding the above two 2-dimensional functions together.
\begin{equation}\label{eq2}
\footnotesize
\epsilon(x,y)\cdot(\frac{p\cdot q}{(M_{R}^{2}-x)^{2}+M_{R}^{2}\cdot \Gamma^{2}}+
\frac{p\cdot q}{(M_{R}^{2}-y)^{2}+M_{R}^{2}\cdot \Gamma^{2}})\otimes\sigma(x,y),
\end{equation}

The resolution for distributions $M^2(\pip\psip)$ and $M^2(\pim\psip)$ are shown in Fig.~\ref{resolution}
for c.m.~energies $\sqrt{s}=$4.416, 4.358, 4.258 and 4.226~GeV. It can be parameterised by a double-Gaussian
with the parameters listed in Table~\ref{tablegau}. The distributions of
2-dimensional efficiency function are shown in Fig.~\ref{efficiency}.

\begin{figure*}[htbp]
\begin{center}
\begin{overpic}[width=4.8cm,height=4.0cm,angle=0]{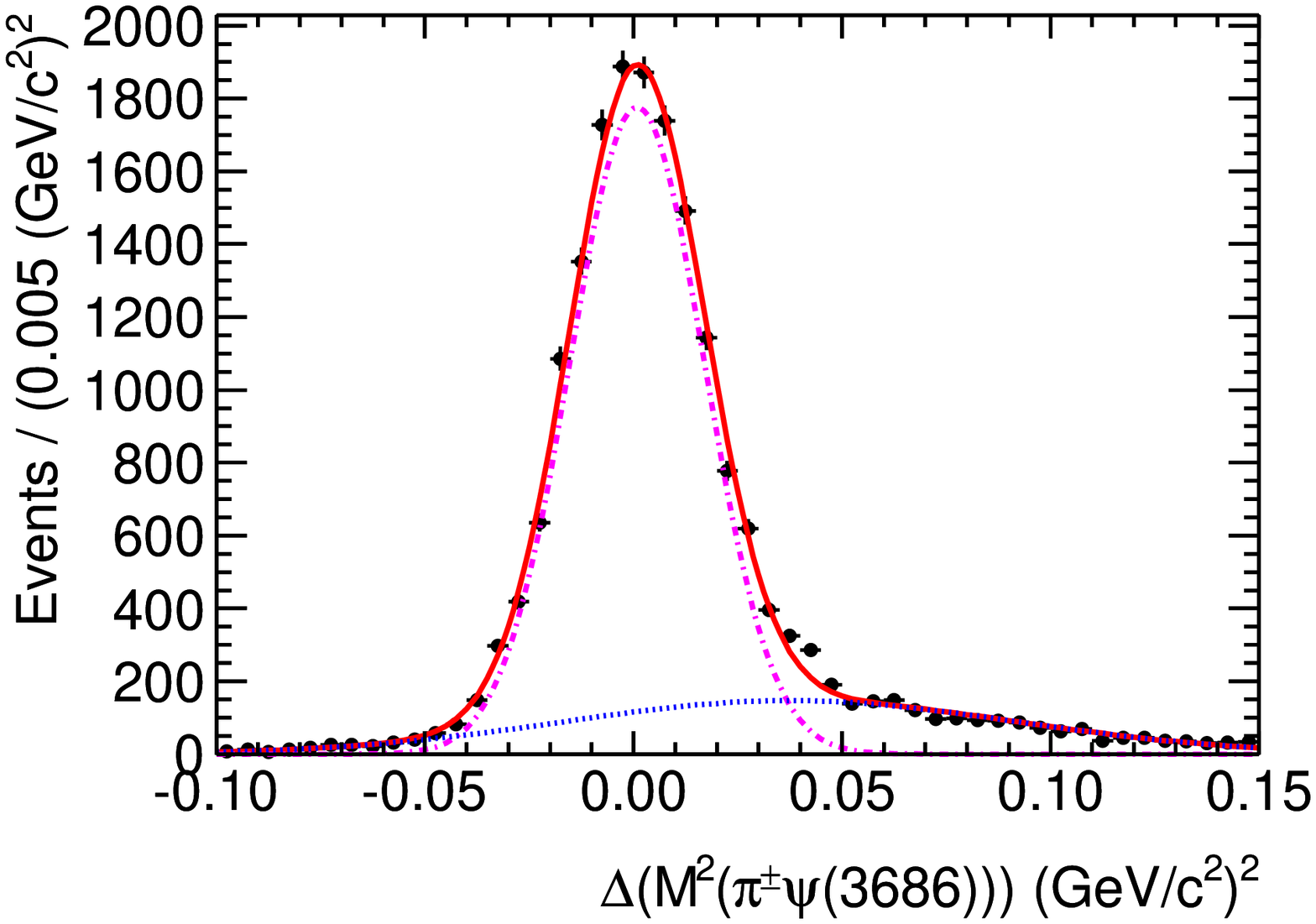}
\put(20,55){ (a)}
\end{overpic}
\begin{overpic}[width=4.8cm,height=4.0cm,angle=0]{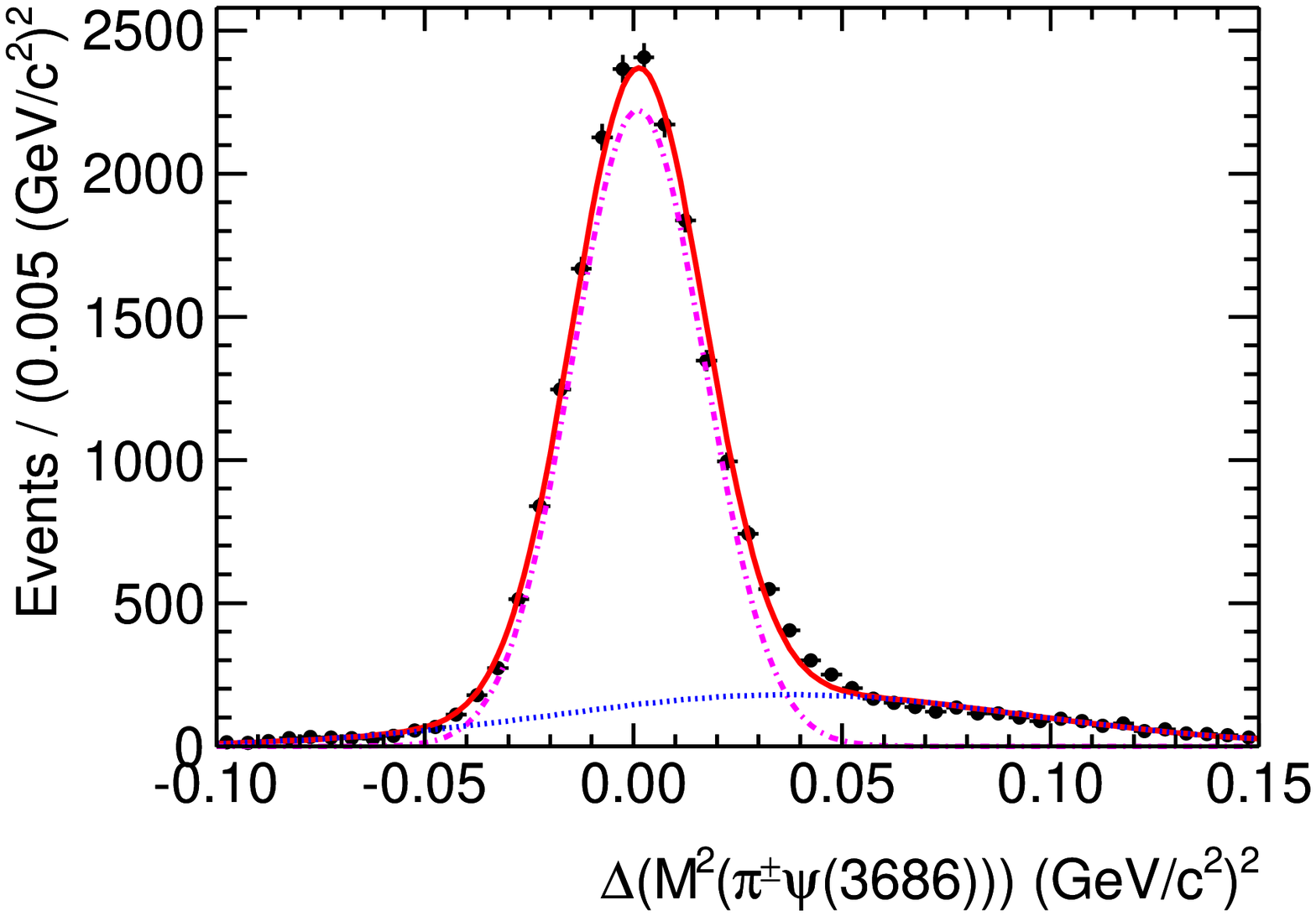}
\put(20,55){ (b)}
\end{overpic}
\begin{overpic}[width=4.8cm,height=4.0cm,angle=0]{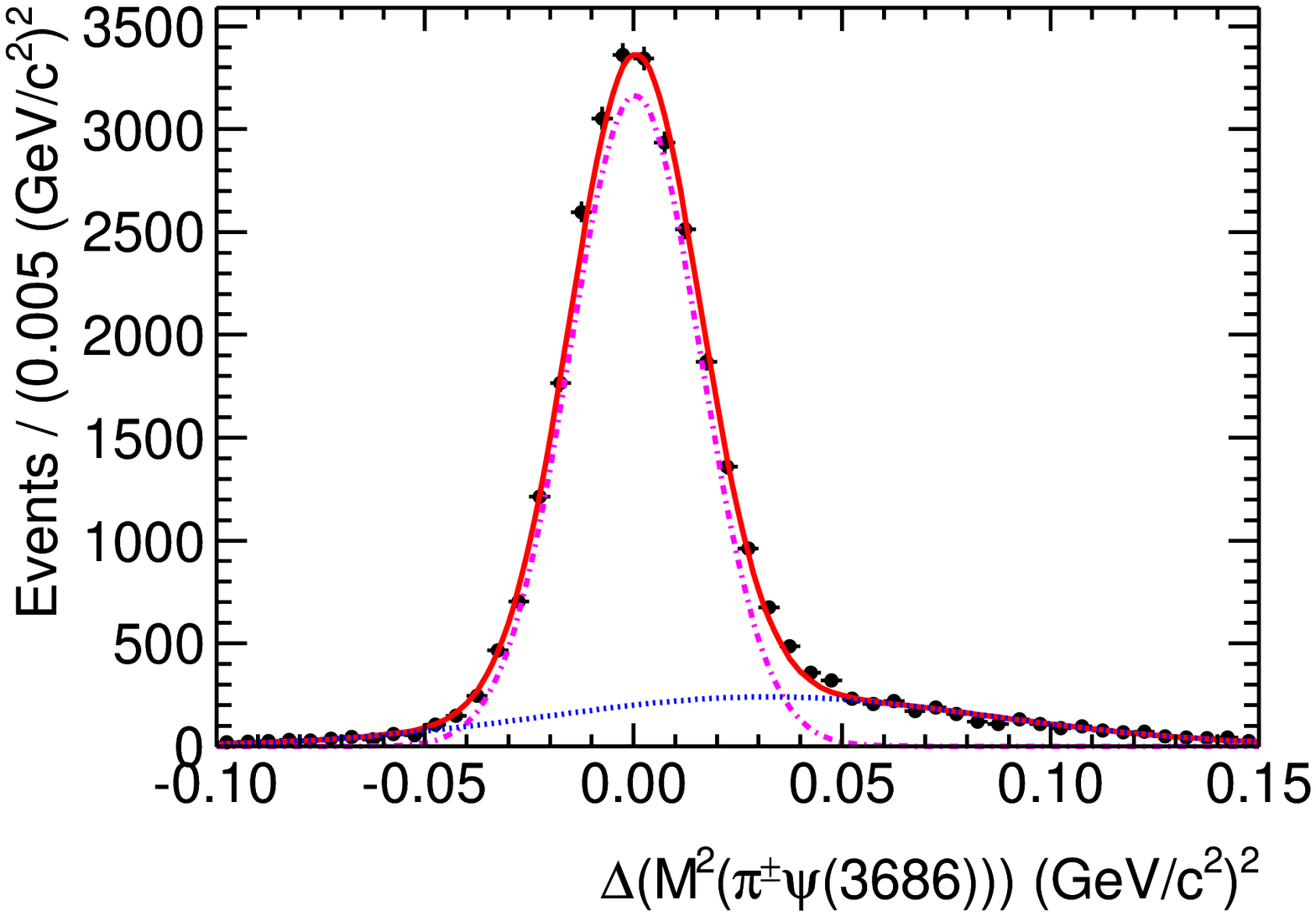}
\put(20,55){ (c)}
\end{overpic}\\
\begin{overpic}[width=4.8cm,height=4.0cm,angle=0]{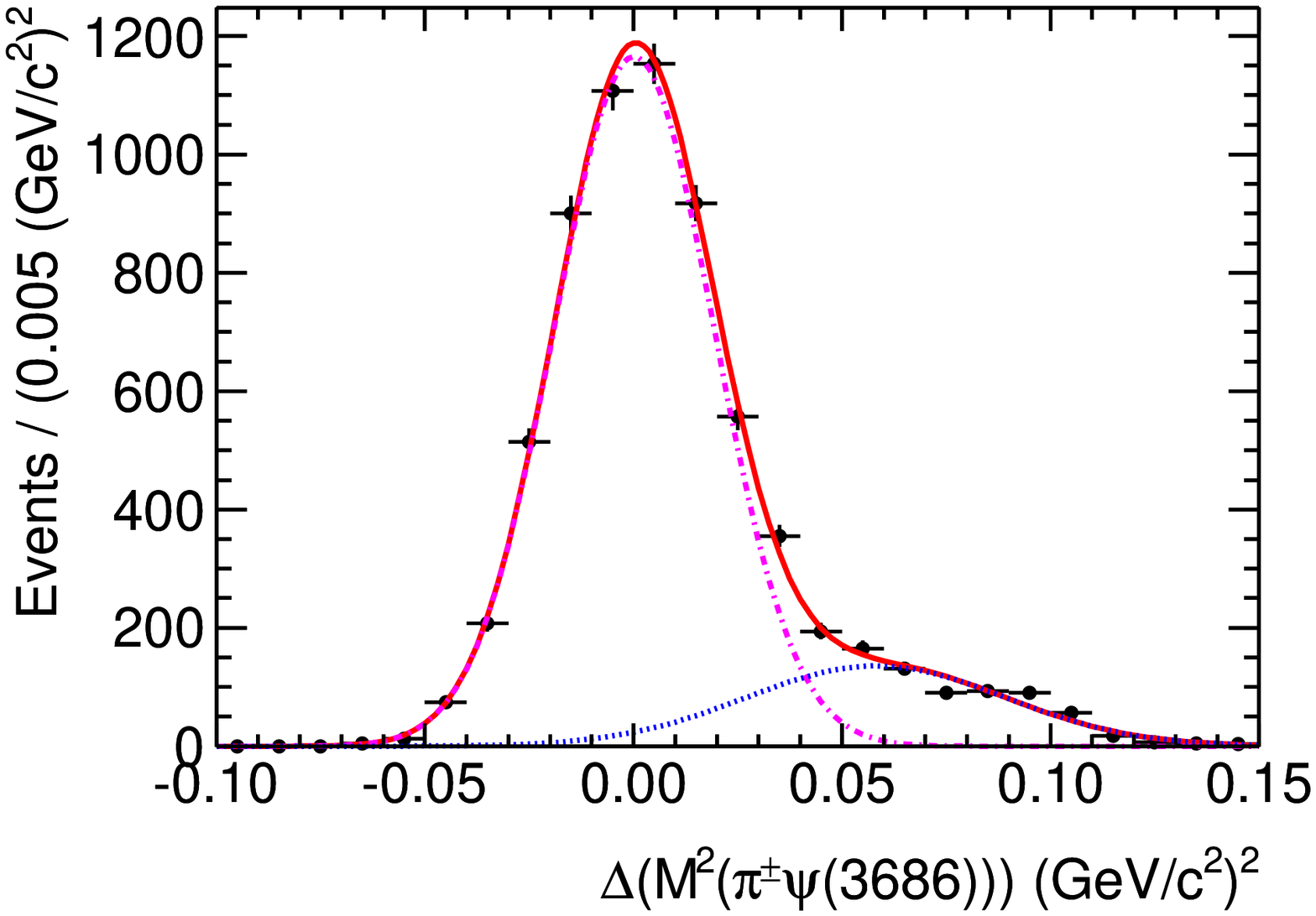}
\end{overpic}
\begin{overpic}[width=4.8cm,height=4.0cm,angle=0]{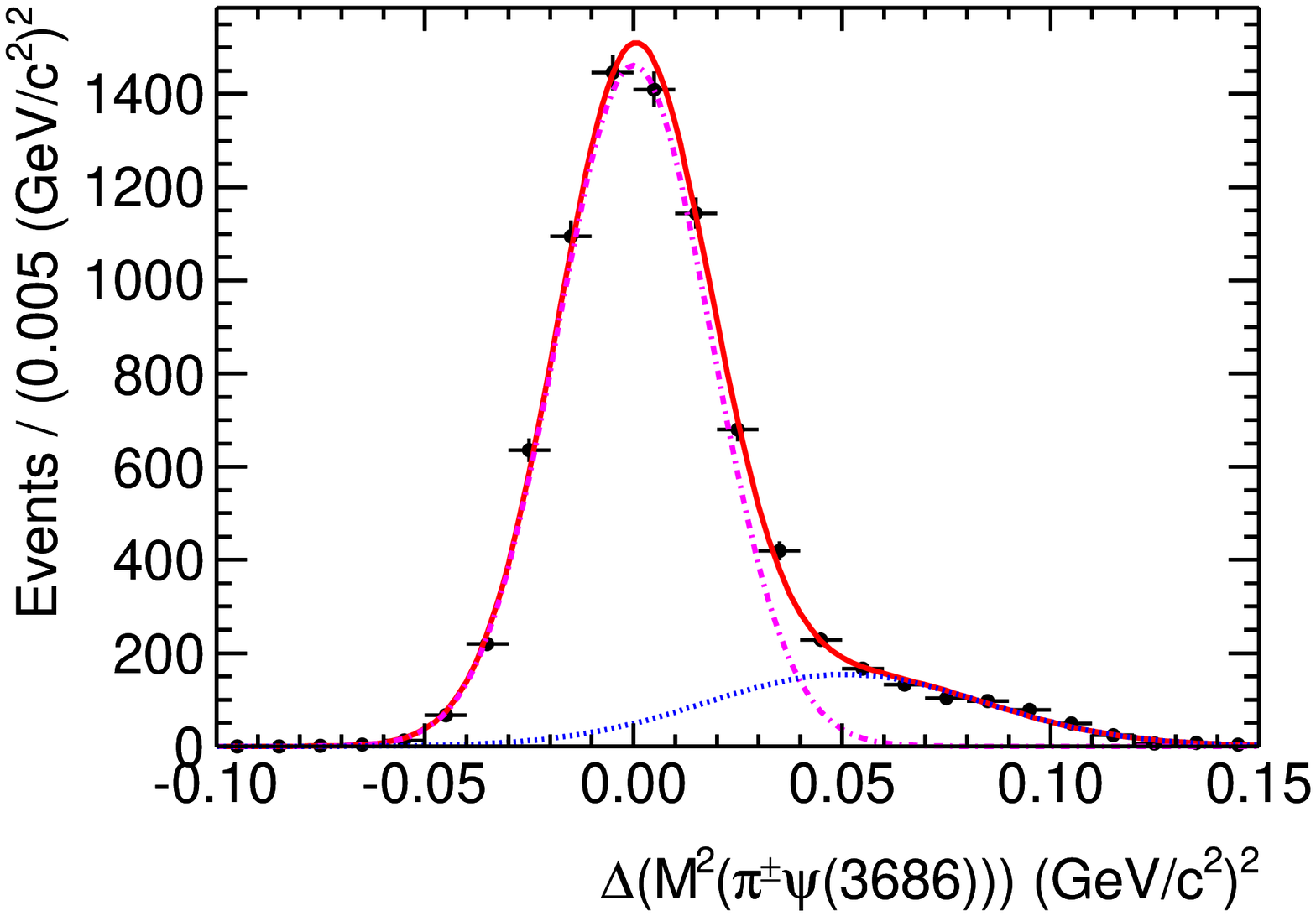}
\end{overpic}
\begin{overpic}[width=4.8cm,height=4.0cm,angle=0]{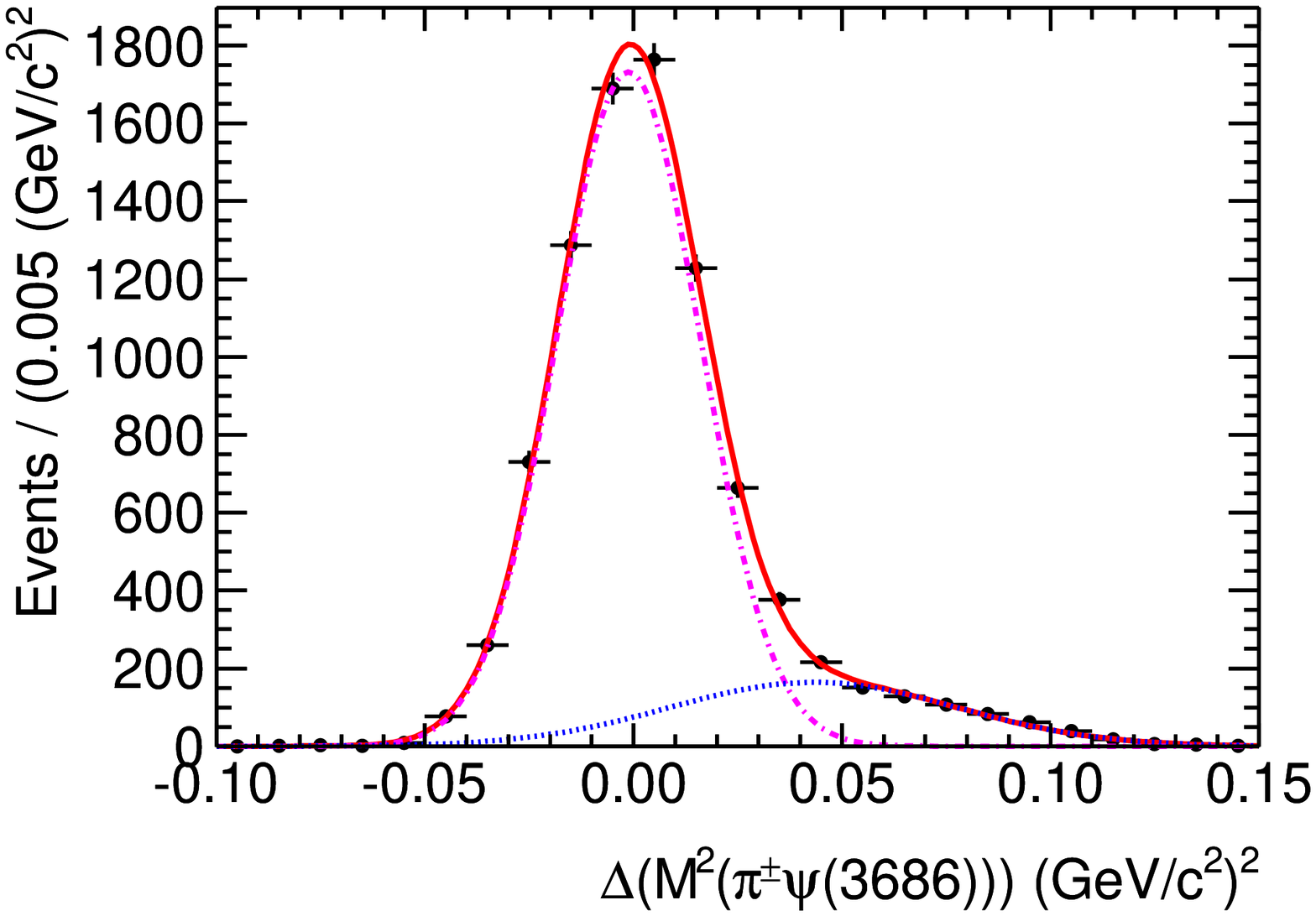}
\end{overpic}
\parbox[1cm]{16cm} {
\caption{Resolution function for $M^2(\pi^{\pm}\psip)$ at (a) 4.416, (b) 4.358, (c) 4.258
and (d) 4.226~GeV. The upper plots for mode I and bottom for mode II.
Dots with error bars are data, the red solid line are fit results,
dot-dashed line in pink denotes contribution from Gaussian function one,
and the dashed line in blue denotes contribution from Gaussian function two.
}
\label{resolution}}
\end{center}
\end{figure*}

\begin{table*}[htbp]
\setlength{\belowcaptionskip}{-0.1cm}
  \caption{Parameters of the resolution for $M^2(\pi^{\pm}\psip)$,
  $M$ and $\sigma$ are the mean and resolution of Gaussian function, respectively.
  The subscript $1$ or $2$ denote Gaussian function one
  or two.
  $fra$ is the fraction of Gaussian function one.
  }\label{tablegau}
  \begin{center}
  \begin{tabular}{ c | c | c|c|c|c}
  \hline
  \hline
  ~~~~$\sqrt{s}$~(GeV)~~~ & ~~$M_{1}$ (GeV/c$^{2}$)$^{2}$~~ & ~~$\sigma_{1}$ (GeV/c$^{2}$)$^{2}$~~ & ~~$M_{2}$ (GeV/c$^{2}$)$^{2}$~~ & ~~$\sigma_{2}$ (GeV/c$^{2}$)$^{2}$~~  & $fra$ \\
  \hline
  4.416~(Mode I) &  $0.00088\pm0.00017$ & $0.01611\pm0.00016$ & $0.0377\pm0.0014$ & $0.0543\pm0.0010$ & $0.7862\pm0.0064$\\
  4.416~(Mode II) & $0.00014\pm0.00047$ & $0.01918\pm0.00031$ & $0.0581\pm0.0046$ & $0.0311\pm0.0023$ &
  $0.841\pm0.019$ \\
  4.358~(Mode I) & $0.00114\pm0.00015$ & $0.01585\pm0.00014$ & $0.0376\pm0.0013$ & $0.0567\pm0.0010$ & $0.7809\pm0.0057$ \\
  4.358~(Mode II) & $0.00011\pm0.00037$ & $0.01839\pm0.00029$ & $0.0508\pm0.0049$ & $0.0335\pm0.0021$ & $0.839\pm0.020$ \\
  4.258~(Mode I) & $0.00037\pm0.00012$    & $0.01562\pm0.00011$ & $0.0326\pm0.0010$ & $0.05346\pm0.00079$ & $0.7974\pm0.0048$ \\
  4.258~(Mode II)& $-0.001167\pm0.00026$ & $0.01724\pm0.00023$ & $0.0432\pm0.0032$ & $0.0344\pm0.0013$ & $0.841\pm0.014$ \\
  \hline \hline
  \end{tabular}
  \end{center}
  \end{table*}

\begin{figure*}[htbp]
\begin{center}
\begin{overpic}[width=4.8cm,height=4.0cm,angle=0]{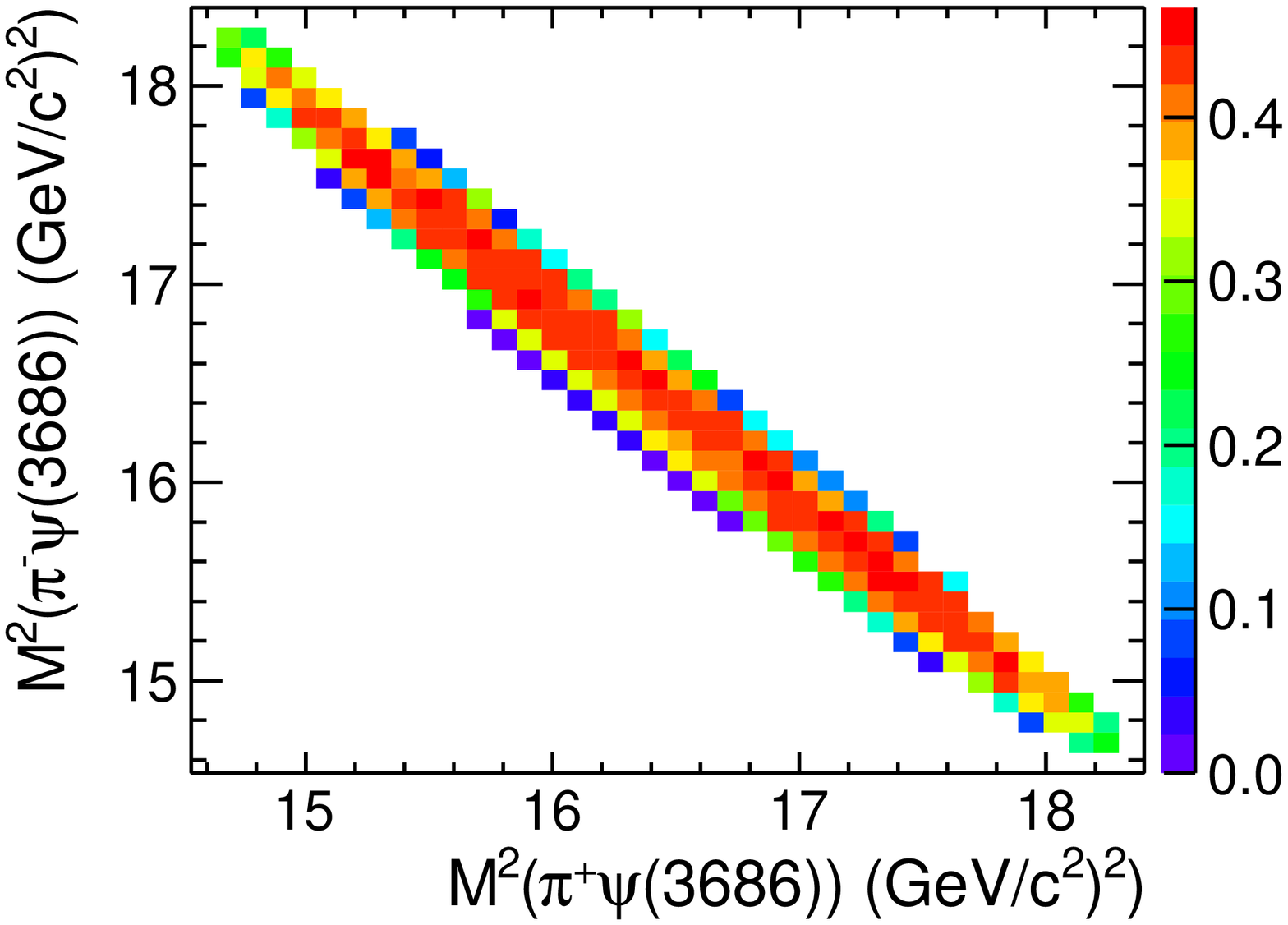}
\put(20,45){ (a)}
\end{overpic}
\begin{overpic}[width=4.8cm,height=4.0cm,angle=0]{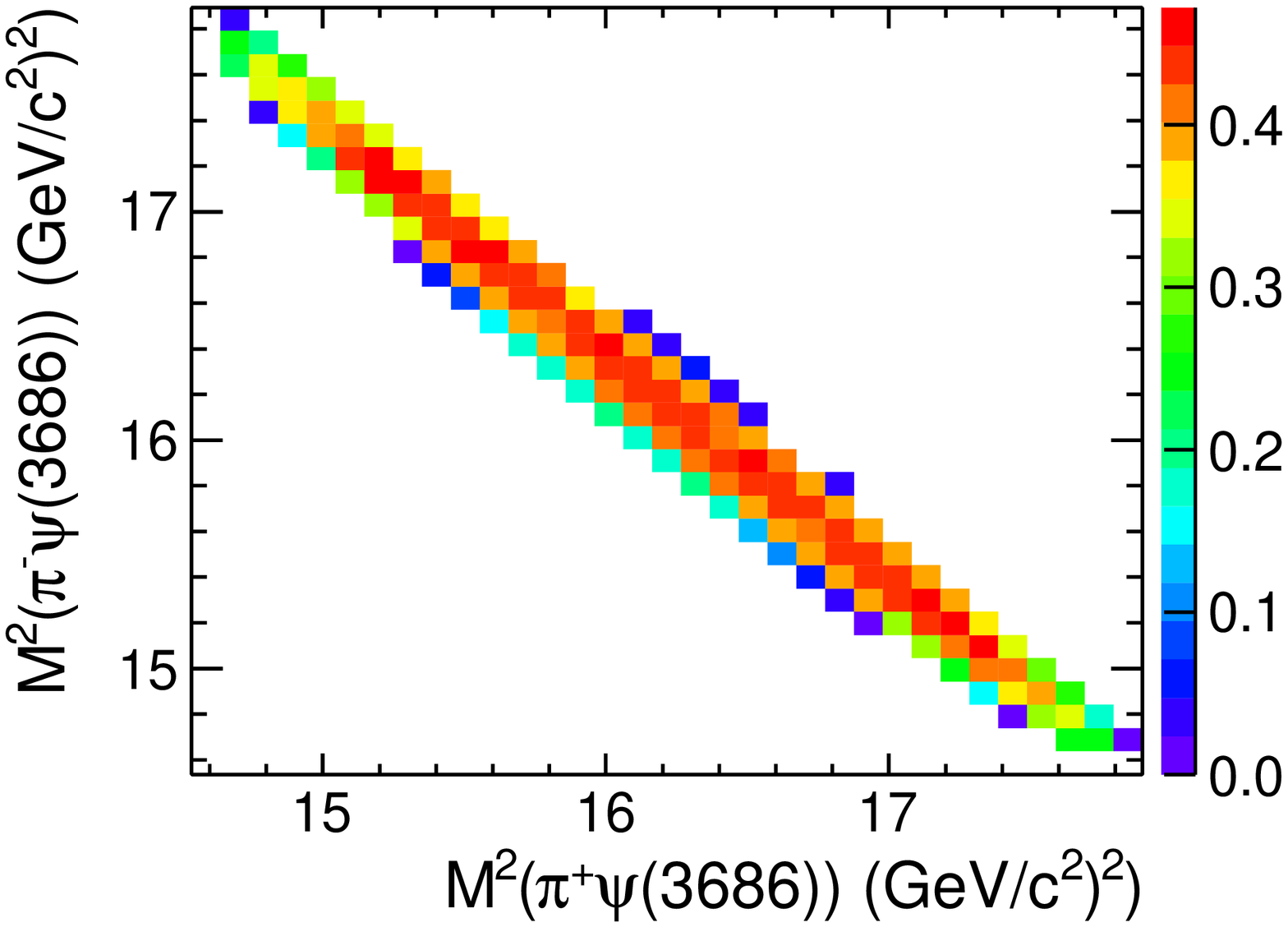}
\put(20,45){ (b)}
\end{overpic}
\begin{overpic}[width=4.8cm,height=4.0cm,angle=0]{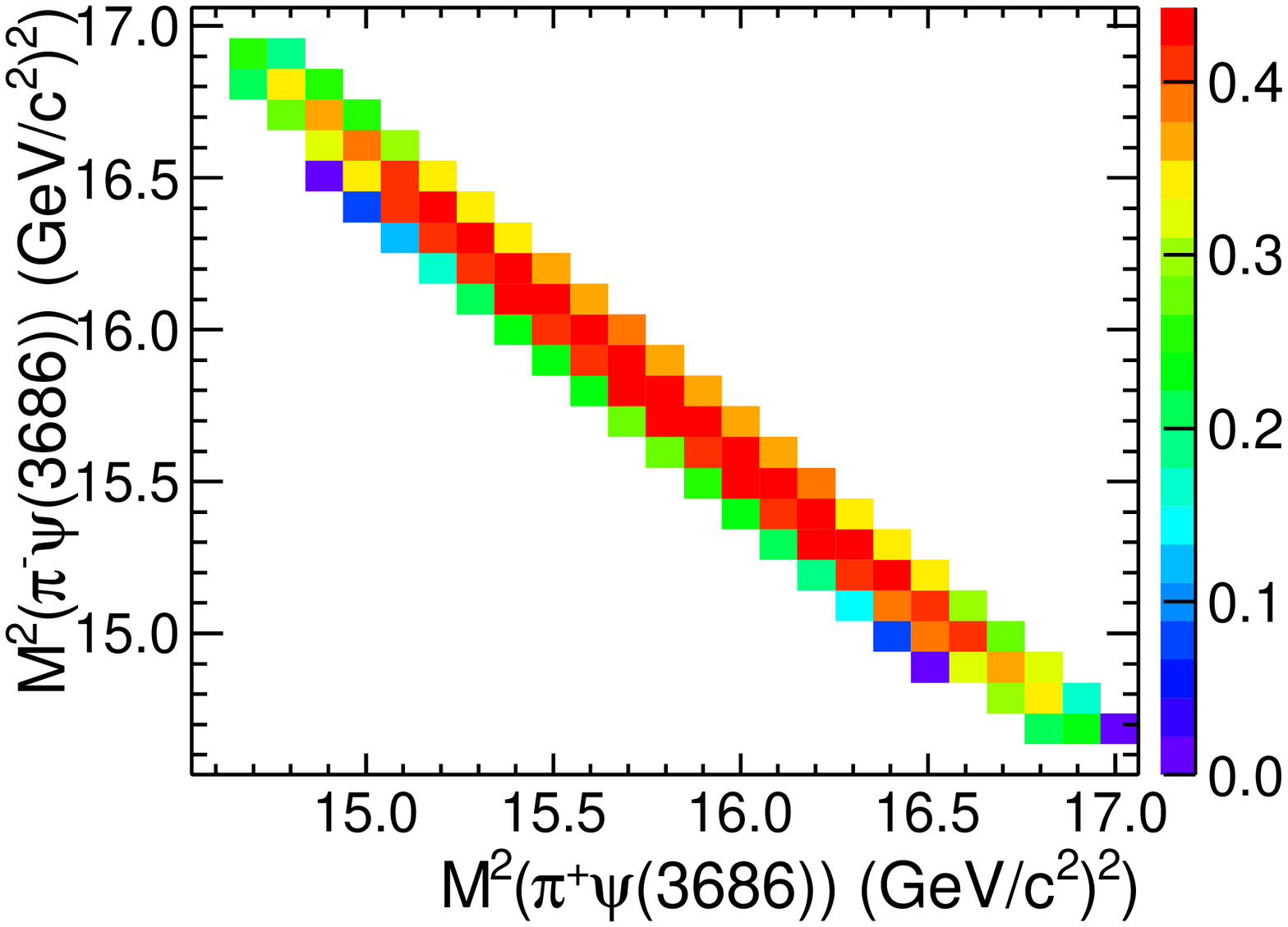}
\put(20,45){ (c)}
\end{overpic}\\
\begin{overpic}[width=4.8cm,height=4.0cm,angle=0]{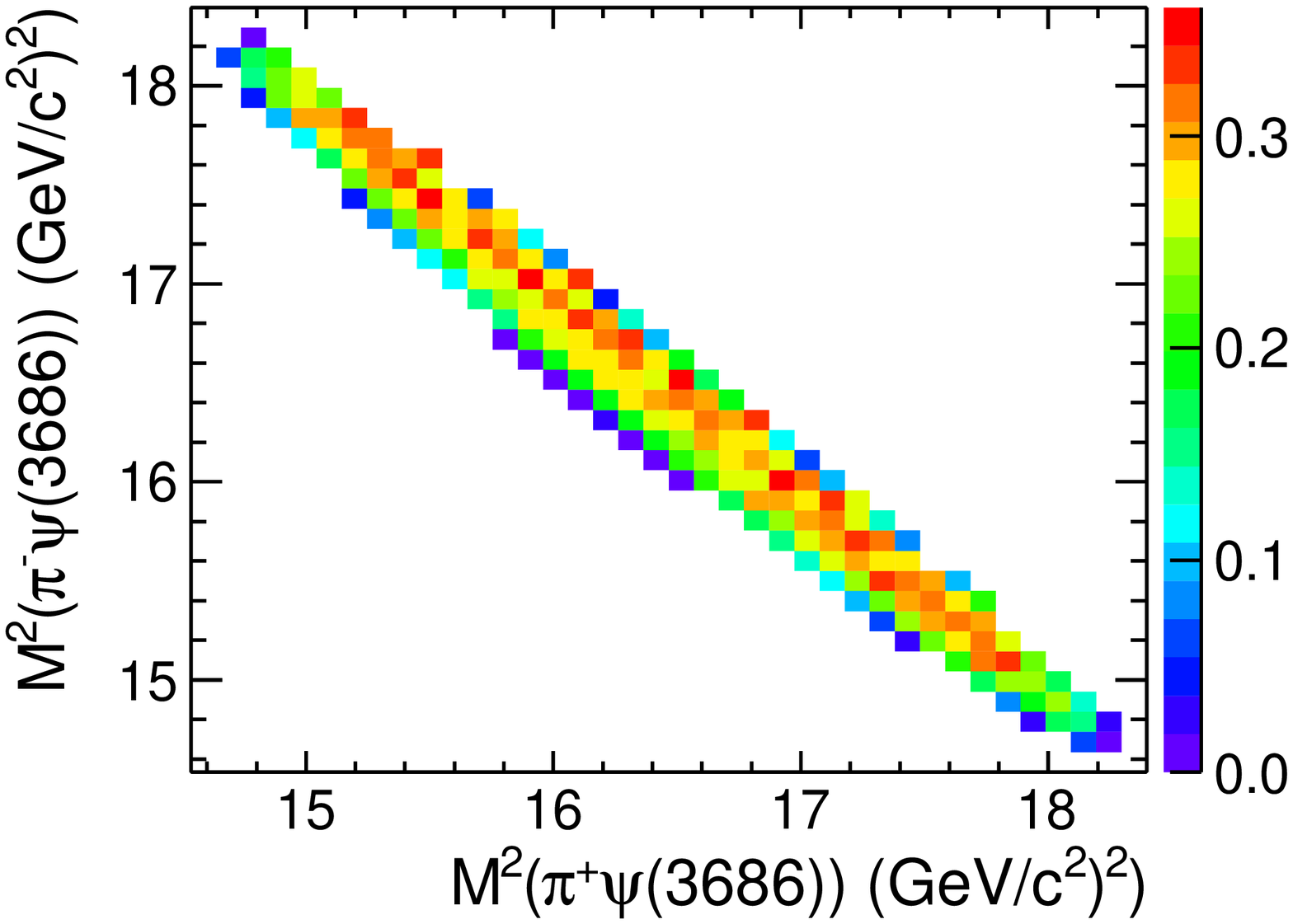}
\end{overpic}
\begin{overpic}[width=4.8cm,height=4.0cm,angle=0]{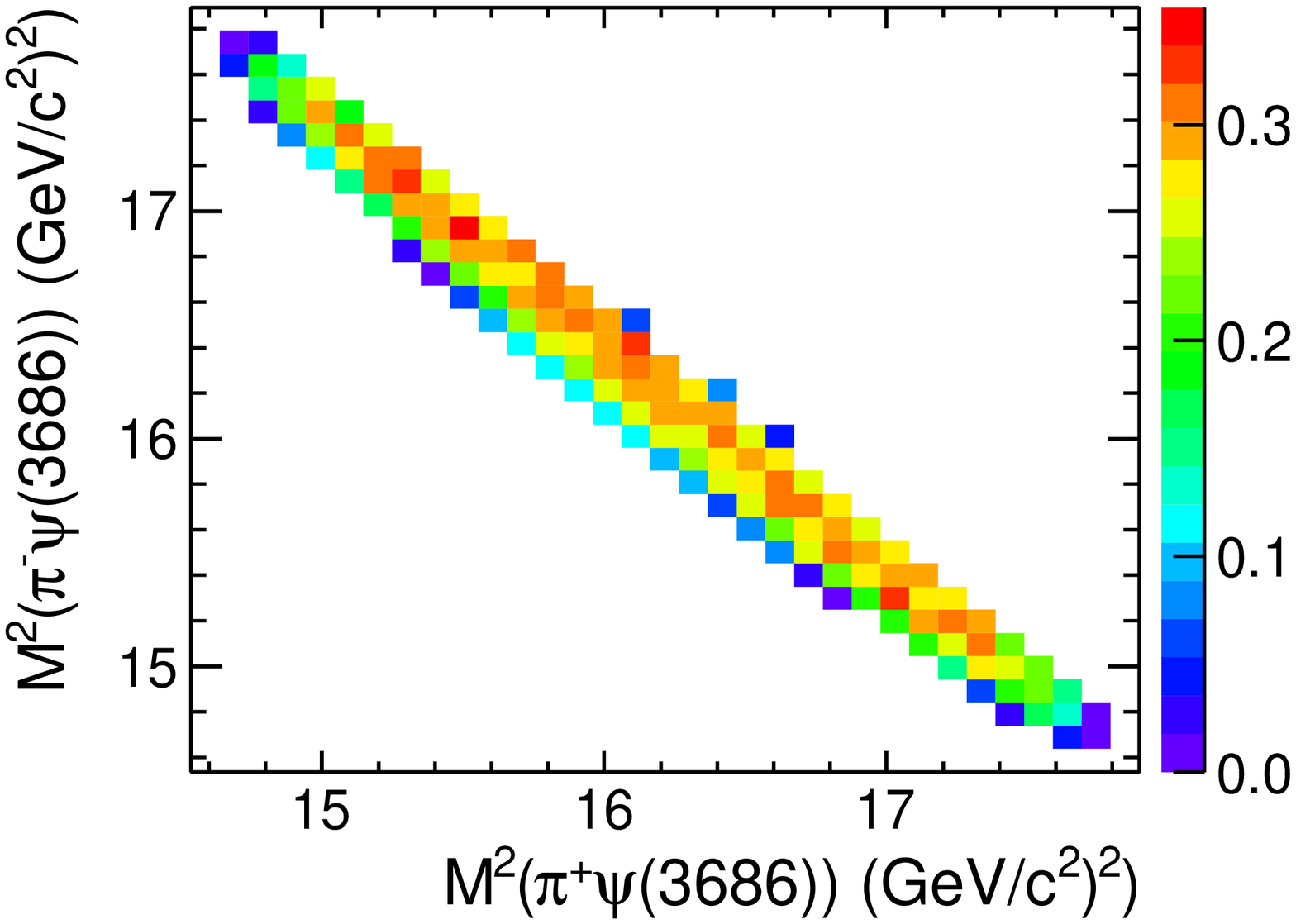}
\end{overpic}
\begin{overpic}[width=4.8cm,height=4.0cm,angle=0]{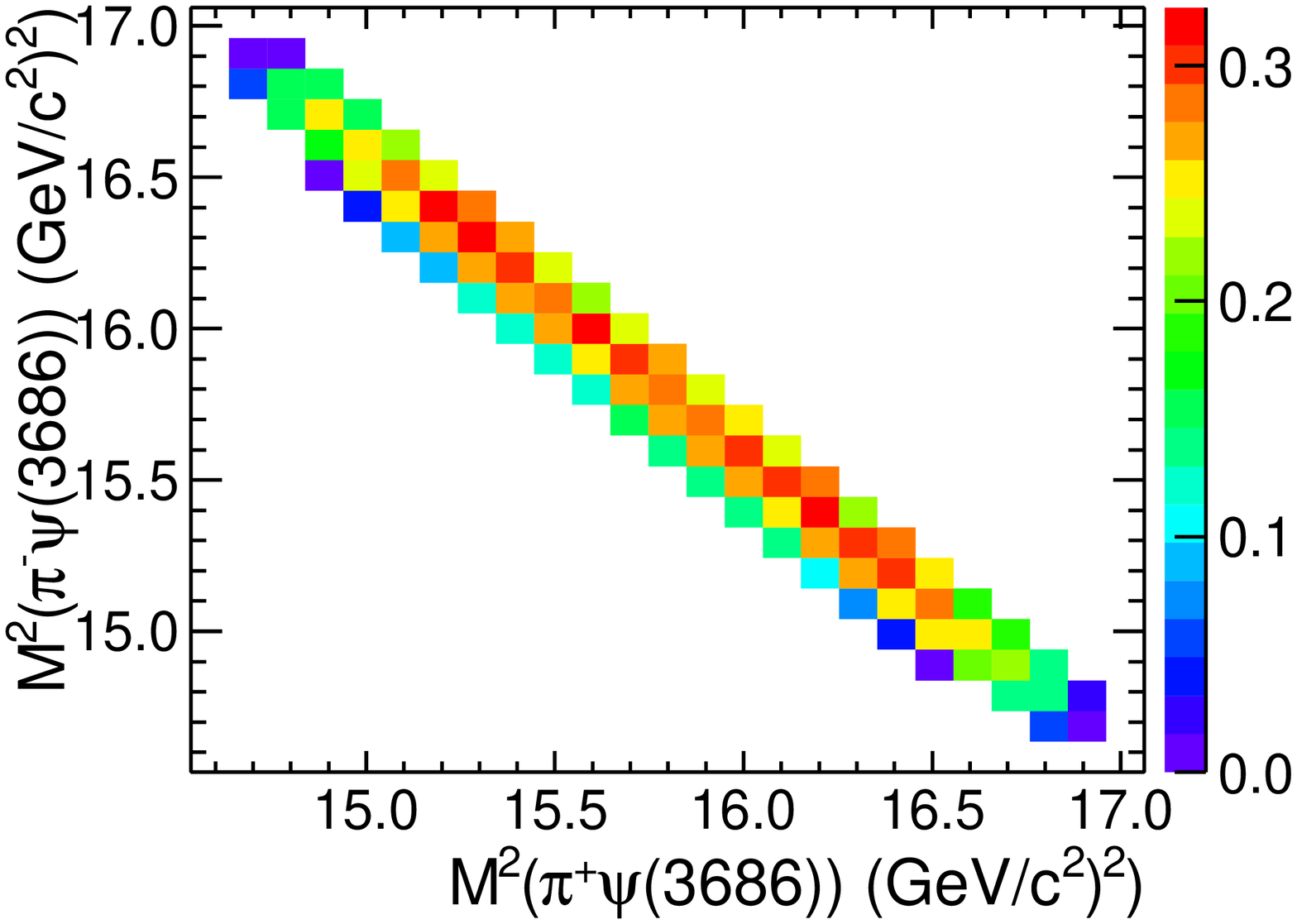}
\end{overpic}
\vskip -0.3cm
\parbox[1cm]{16cm} {
\caption{ The 2-dimensional efficiency curve $\epsilon(x,y)$ for $M^2(\pip\psip)$ versus $M^2(\pim\psip)$
at (a) 4.416, (b) 4.358, (c) 4.258 and (d) 4.226~GeV. The upper plots for mode I and bottom for mode II.
}
\label{efficiency}}
\end{center}
\end{figure*}


\begin{thebibliography}{99}
\bibitem{Y4260BaBar} B. Aubert {\it et al.} [BaBar Collaboration], Phys.\ Rev.\ Lett. {\bf 95}, 142001 (2005);
                     J. P. Lees {\it et al.} [BaBar Collaboration], Phys.\ Rev.\ D {\bf 86}, 051102(R) (2012);
\bibitem{Y4260Cleo}  Q. He {\it et al.} [CLEO Collaboration], Phys.\ Rev.\ D {\bf 74}, 091104(R) (2006);
\bibitem{Y4260Belle} C. Z. Yuan {\it et al.} [Belle Collaboration], Phys.\ Rev.\ Lett. {\bf 99}, 182004 (2007);
                     Z. Q. Liu {\it et al.} [Belle Collaboration], Phys.\ Rev.\ Lett. {\bf 110}, 252002 (2013).
\bibitem{Y4360BaBar} B. Aubert {\it et al.} [BaBar Collaboration], Phys.\ Rev.\ Lett. {\bf 98}, 212001 (2007);
                     Phys.\ Rev.\ D {\bf 89}, 111103 (2014).
\bibitem{Y4360Belle} X. L. Wang {\it et al.} [Belle Collaboration], Phys.\ Rev.\ Lett. {\bf 99}, 142002 (2007);
                     Phys.\ Rev.\ D {\bf 91}, 112007 (2015).

\bibitem{Y4360hybrid}  F. E. Close and P. R. Page, Phys.\ Lett.\ B {\bf628}, 215 (2005);
\bibitem{Y4360tetra}   L. Maiani, V. Riquer, F. Piccinini and A. D. Polosa, Phys.\ Rev.\ D {\bf 72},
                       031502 (2005);
\bibitem{Y4360molecule} M.~B.~Voloshin and L.~B.~Okun, JETP Lett.\ {\bf 23}, 333 (1976);
                    A.~De Rujula, H.~Georgi and S.~L.~Glashow, Phys.\ Rev.\ Lett.\ {\bf 38}, 317 (1977);
X. Liu, X. Q. Zeng and X. Q. Li, Phys.\ Rev.\ D {\bf 72}, 054023 (2005);
\bibitem{pipijpsi} M.~Ablikim {\it et al.} [BESIII Collaboration], Phys.\ Rev.\ Lett.\ {\bf 118}, 092001 (2017).
\bibitem{pipihc} M.~Ablikim {\it et al.} [BESIII Collaboration], Phys.\ Rev.\ Lett.\ {\bf 118}, 092002 (2017).
\bibitem{ZC3900BES}  M. Ablikim {\it et al.} [BESIII Collaboration], Phys.\ Rev.\ Lett. {\bf 110}, 252001 (2013).
\bibitem{ZC3900CLEO} T. Xiao {\it et al.}, Phys.\ Lett.\ B {\bf 727}, 366 (2013).
\bibitem{ZC4020BES} M. Ablikim {\it et al.} [BESIII Collaboration], Phys.\ Rev.\ Lett. {\bf 111}, 242001 (2013).
\bibitem{ZC3885BES} M. Ablikim {\it et al.} [BESIII Collaboration], Phys.\ Rev.\ Lett. {\bf 112}, 022001 (2014).
\bibitem{ZC3885BES2} M.~Ablikim {\it et al.} [BESIII Collaboration], Phys.\ Rev.\ D {\bf 92}, 092006 (2015).
\bibitem{ZC4025BES} M. Ablikim {\it et al.} [BESIII Collaboration], Phys.\ Rev.\ Lett. {\bf 112}, 132001 (2014).
\bibitem{zb} A. Bondar {\it et al.} [Belle Collaboration], Phys.\ Rev.\ Lett. {\bf 108},122001(2012).
\bibitem{cme} M.~Ablikim {\it et al.} [BESIII Collaboration], Chin.\ Phys.\ C {\bf 40}, 063001 (2016).
\bibitem{luminosity} M. Ablikim {\it et al.} [BESIII Collaboration], Chin.\ Phys.\ C {\bf 39}, 093001 (2015).
\bibitem{besint} M. Ablikim {\it et al.} [BESIII Collaboration], Nucl.\ Instrum.\ Meth.\ A {\bf 614}, 3 (2010).
\bibitem{geant4} S. Agostinelli {\it et al.} [{\sc Geant4} Collaboration], Nucl.\ Instrum.\ Meth. A {\bf 506}, 250 (2003).
\bibitem{Deng} Z.~Y.~Deng $et~al.$, HEP\&NP {\bf 30}, 371 (2006).
\bibitem{kkmc} S. Jadach, B. F. L. Ward, and Z. Was, Phys.\ Rev.\ D {\bf 63}, 113009 (2001).
\bibitem{photos} E. Barberio and Z. Was, Comput.\ Phys.\ Commum.\ {\bf 79}, 219 (1994).
\bibitem{jpipi} T.~Mannel and R.~Urech, Z.\ Phys.\ C {\bf 73}, 541 (1997);
S.~Chakravarty and P.~Ko, Phys.\ Rev.\ D {\bf 48}, 1205 (1993);
S.~Chakravarty, S.~M.~Kim and P.~Ko, Phys.\ Rev.\ D {\bf 50}, 389 (1994).
\bibitem{evtgen} R. G. Ping, Chin.\ Phys.\ C {\bf 32}, 599 (2008);
                D. J. Lange, Nucl.\ Instrum.\ Meth.\ A {\bf 462}, 152 (2001).
\bibitem{pdg} K.~A.~Olive {\it et al.} [Particle Data Group], Chin.\ Phys.\ C, {\bf 38}, 090001 (2014).
\bibitem{lundcharm} J. C. Chen, G. S. Huang, X. R. Qi, D. H. Zhang and Y. S. Zhu, Phys. Rev. D {\bf 62}, 034003 (2000).
\bibitem{uplimit}W.~A.~Rolke {\it et al.} Nucl.\ Instrm.\ Meth.\ A {\bf 551}, 493 (2005).
\bibitem{trkuncertainty} M.~Ablikim {\it et al.} [BESIII Collaboration], Phys.\ Rev.\ Lett. {\bf 112}, 022001 (2014);
Phys.\ Rev.\ D {\bf 83}, 112005 (2015).
\bibitem{phouncertainty} M.~Ablikim {\it et al.} [BESIII Collaboration], Phys.\ Rev.\ D {\bf 81}, 052005 (2010).
\bibitem{kinematic} M.~Ablikim {\it et al.} [BESIII Collaboration], Phys.\ Rev.\ D {\bf 87}, 012002 (2013).
\bibitem{vacuum} E.~A.~Kuraev and V.~S.~Fadin, Sov.\ J.\ Nucl.\ Phys. {\bf 41}, 466 (1985) [Yad.\ Fiz.\ {\bf 41}, 733 (1985)].
\bibitem{combine1} G.~D'Agostini, Nucl.\ Instrum.\ Meth.\ A {\bf 346}, 306 (1994).
\bibitem{combine2} M.~Ablikim {\it et al.} [BESIII Collaboration], Phys.\ Rev.\ D {\bf 89}, 074039 (2014).
\bibitem{espread}E.~V.~Abakumova {\it et al.} Nucl.\ Instrum.\ Meth.\ A {\bf 659}, 21 (2011).
\bibitem{omegachicj}M.~Ablikim {\it et al.} [BESIII Collaboration], Phys.\ Rev.\ Lett.\ {\bf 114}, 092003 (2015).
\bibitem{piDD} Chang-Zheng Yuan for the BESIII Collaboration, "$e^{+}e^{-}$ annihilation cross section measurements at BESIII", talk at "the 10th workshop of the France China Particle Physics Laboratory", \url{http://indico.ihep.ac.cn/event/6651/timetable/#20170329.detailed}, March 27-30, 2017, Tsinghua University, Beijing, China.
\bibitem{zoubs}M.~B.~Voloshin, JETP Lett.\ {\bf 37}, 69 (1983);
V.~V.Anisovich, D.~V.~Bugg, A.~V.~Sarantsev and B.~S.~Zou, Phys.\ Rev.\ D {\bf 51}, R4619 (1995).\\
\bibitem{combinefit} X.~Y.~Gao, C.~P.~Shen and C.~Z.~Yuan, Phys.\ Rev.\ D {\bf 95}, 092007 (2017).\\
\bibitem{etajpsi} M.~Ablikim {\it et al.} [BESIII Collaboration], Phys.\ Rev.\ D {\bf 91}, 112005 (2015).

\end{thebibliography}
\end{document}